\documentclass[apj]{emulateapj}
\usepackage[colorlinks=true,linkcolor=blue,citecolor=red]{hyperref}
\usepackage{bm}
\usepackage{graphicx}
\usepackage{epsf}
\usepackage{graphics}
\usepackage{amsmath}

\def\beq{\begin{equation}}
\def\eeq{\end{equation}}
\def\bey{\begin{eqnarray}}
\def\eey{\end{eqnarray}}

\def\mpc{\, h^{-1}{\rm {Mpc}}}

\def\Msun{{\rm M_\odot}}
\def\msun{\, h^{-1}{\rm M_\odot}}
\def\msunt{\, h^{-2}{\rm M_\odot}}

\def\gs{\mathrel{\raise1.16pt\hbox{$>$}\kern-7.0pt
\lower3.06pt\hbox{{$\scriptstyle \sim$}}}}
\def\ls{\mathrel{\raise1.16pt\hbox{$<$}\kern-7.0pt
\lower3.06pt\hbox{{$\scriptstyle \sim$}}}}
\def\gtsima{\, {\buildrel > \over \sim} \,}
\def\ltsima{\, {\buildrel < \over \sim} \,}
\def\prosima{\, {\buildrel \propto \over \sim} \,}
\def\gsim{\lower.5ex\hbox{\gtsima}}
\def\lsim{\lower.5ex\hbox{\ltsima}}
\def\simgt{\lower.5ex\hbox{\gtsima}}
\def\simlt{\lower.5ex\hbox{\ltsima}}
\def\simpr{\lower.5ex\hbox{\prosima}}

\shorttitle{ELUCID: Galaxy Quenching and its Relation to Environment}
\shortauthors{Wang H.Y. et al.}

\begin{document}

\title {ELUCID IV: Galaxy Quenching and its Relation to Halo Mass, Environment, and Assembly Bias}
\author{Huiyuan Wang\altaffilmark{1,2}, H.J. Mo\altaffilmark{3,4}, Sihan Chen\altaffilmark{1}, Yang Yang\altaffilmark{1}, Xiaohu Yang\altaffilmark{5,6}, Enci Wang\altaffilmark{1,2}, Frank C. van den Bosch\altaffilmark{7}, Yipeng Jing\altaffilmark{5,6}, Xi Kang\altaffilmark{8}, Weipeng Lin\altaffilmark{9}, S.H. Lim\altaffilmark{3}, Shuiyao Huang\altaffilmark{3}, Yi Lu\altaffilmark{11}, Shijie Li\altaffilmark{5,6}, Weiguang Cui\altaffilmark{10}, Youcai Zhang\altaffilmark{11},  Dylan Tweed\altaffilmark{5,6}, Chengliang Wei\altaffilmark{8}, Guoliang Li\altaffilmark{8} and Feng Shi\altaffilmark{11}}

\altaffiltext{1}{Key Laboratory for Research in Galaxies and Cosmology, Department of Astronomy, University of Science and
Technology of China, Hefei, Anhui 230026, China; whywang@mail.ustc.edu.cn}
\altaffiltext{2}{School of Astronomy and Space Science, University of Science and Technology of China, Hefei 230026, China}
\altaffiltext{3}{Department of Astronomy, University of Massachusetts, Amherst MA 01003-9305, USA}
\altaffiltext{4}{Astronomy Department and Center for Astrophysics, Tsinghua University, Beijing 10084, China}
\altaffiltext{5}{Department of Astronomy, Shanghai Jiao Tong University, Shanghai 200240, China}
\altaffiltext{6}{IFSA Collaborative Innovation Center, Shanghai Jiao Tong University, Shanghai 200240, China}
\altaffiltext{7}{Department of Astronomy, Yale University, P.O. Box 208101, New Haven, CT 06520-8101, USA}
\altaffiltext{8}{Purple Mountain Observatory, the Partner Group of MPI f\"ur Astronomie,  2 West beijing Road, Nanjing 210008, China}
\altaffiltext{9}{School of Physics and Astronomy, Sun Yat-Sen University, Guangzhou 510275, China}
\altaffiltext{10}{Departamento de F\'isica Te\'orica, M\'odulo 15, Facultad de Ciencias, Universidad Aut\'onoma de Madrid, E-28049 Madrid, Spain}
\altaffiltext{11}{Shanghai Astronomical Observatory, Nandan Road 80, Shanghai 200030, China}
\begin{abstract}
We examine the quenched fraction of central and satellite galaxies as a function
of galaxy stellar mass, halo mass, and the matter density of their large scale environment.
Matter densities are inferred from our ELUCID simulation, a constrained simulation
of local Universe sampled by SDSS, while halo masses and central/satellite classification are
taken from the galaxy group catalog of Yang et al. The quenched fraction for the total population
increases systematically with the three quantities. We find that the `environmental quenching efficiency',
which quantifies the quenched fraction as function of halo mass, is independent of stellar mass.
And this independence is the origin of the stellar mass-independence of density-based quenching efficiency,
found in previous studies. Considering centrals and satellites separately, we find that the two
populations follow similar correlations of quenching efficiency with halo mass and stellar mass,
suggesting that they have experienced similar quenching processes in their host halo.
We demonstrate that satellite quenching alone cannot account for the environmental quenching efficiency
of the total galaxy population and the difference between the two populations found previously mainly
arises from the fact that centrals and satellites of the same stellar mass reside, on average, in halos
of different mass. After removing these halo-mass and stellar-mass effects, there
remains a weak, but significant, residual dependence on environmental density, which is eliminated when
halo assembly bias is taken into account. Our results therefore indicate that halo mass is
the prime environmental parameter that regulates the quenching of both centrals and satellites.
\end{abstract}

\keywords{dark matter - large-scale structure of the universe -
galaxies: halos - methods: statistical}

\section{Introduction}
\label{sec_intro}

In the low-redshift universe, galaxies are observed to exhibit bimodal distributions
in their colors and specific star formation rates. According to these bimodal
distributions, galaxies can be divided into two distinctive sequences,
a red sequence of galaxies quenched in star formation,  and a blue sequence
of star-forming galaxies (e.g. Strateva et al. 2001; Blanton et al. 2003; Baldry et al. 2004;
Brinchmann et al. 2004). The bimodal distribution is correlated with galaxy mass,
with the red sequence dominated by massive galaxies, while the blue population
by lower-mass ones. Furthermore, it is also known that the red, quiescent galaxies tend to
reside in high density regions, while the blue star-forming galaxies display an opposite
trend with environments (e.g. Oemler 1974; Dressler 1980; Hogg et al. 2003;
Kauffmann et al. 2004; Baldry et al. 2006; Peng et al. 2010; Zheng et al. 2017). All these suggest that
the mechanisms responsible for quenching star formation in a galaxy
must be related to galaxy mass as well as to the environment.

In the current cold dark matter (CDM) cosmogony, galaxies are assumed to form and evolve
within dark matter halos (e.g. White \& Rees 1978; Mo, van den Bosch \& White 2010).
Lower-mass halos on average form earlier and subsequently merge to form more massive
ones. In this process, the galaxy that forms on the main branch of a halo merging
tree is expected to be the dominant galaxy residing near the center of the halo
(the central galaxy),  while other galaxies that form in the sub-branch progenitor
halos, are expected to orbit the central as satellite galaxies. These two populations
of galaxies are expected to have experienced different processes that quench
their star formation: while supernovae and active galactic nuclei (AGNs) feedbacks,
and shock heating of cold accretion flow may affect both central and satellite galaxies,
processes like ram pressure and tidal stripping are believed to operate only on
satellite galaxies (e.g. White \& Frenk 1991; Kang et al. 2005; Dekel \& Birnboim 2006; Bower et al. 2006;
Croton et al. 2006; Kere{\v s} et al. 2009; Lu et al. 2011; Guo et al. 2011; Vogelsberger et al.2014;
Schaye et al. 2015). These models have successfully reproduced many global properties of observed galaxies.
However, there are still significant discrepancies between model
predictions incorporating these processes and observational data in terms
of the fraction of the quenched population (see e.g. Hirschmann et al. 2014;
Vogelsberger et al. 2014; Henriques et al. 2016),
indicating that our understanding of the quenching processes is still incomplete.

Observing correlations between galaxy properties and different aspects of their environment could
help to distinguish different galaxy formation models. A variety of quantities have been used to describe
the environment around a galaxy (Haas, Schaye \& Jeeson-Daniel 2012). These environmental parameters are
usually designed for different purposes, and an optimal decision has to be made for a specific
question (Leclercq et al. 2016). The traditional environmental parameter is the (projected) number density
of galaxies, which is often adopted in observational studies of galaxy properties on various scales
(e.g. Dressler 1980;  Hogg et al. 2003; Kauffmann et al. 2004; Hirschmann et al. 2014). It can be
directly obtained from the galaxy redshift survey without any additional assumption. The other widely
used parameter is the host halo mass (e.g. Weinmann et al. 2006; Wetzel, Tinker \& Conroy 2012; Woo et al. 2013),
which is closely linked to galaxy formation in the CDM paradigm. Indeed, the halo occupation distribution (HOD)
models (e.g. Jing et al. 1998; Peacock \& Smith 2000; Zheng et al. 2005; Zu \& Mandelbaum 2016), conditional
luminosity function (CLF) models(e.g., Yang et al. 2003;  van den Bosch et al. 2007), abundance
matching models (e.g., Mo et al. 1999; Kravtsov  et al. 2004; Vale \& Ostriker 2006; Behroozi, Conroy, \& Wechsler 2010;
Hearin \& Watson 2013) and halo-based empirical models (e.g., Yang et al. 2013; Lu et al. 2014, 2015; Moster, Naab \& White 2017)
have all used halo masses to link galaxies to dark matter halos.

Using halo mass inferred from galaxy group catalog (Yang et al. 2007) as environmental parameter,
Weinmann et al. (2006) found that the quenched fraction of satellite galaxies is much lower than that in model
predictions and increases strongly with host halo mass (see also Liu et al. 2010;
Wetzel et al. 2012). This has motivated later semi-analytic galaxy
formation models (SAMs) to employ an incremental stripping of hot gas associated with
satellites through ram-pressure and tidal stripping (e.g. Kang \& van den Bosch 2008;
Font et al. 2008; Weinmann et al. 2010; Guo et al. 2011; Henriques et al. 2015). Moreover,
halo mass is also found to have significant impact on the quenching of star
formation in centrals of given galaxy masses (Weinmann et al. 2006; Woo et al. 2013; 2015;
Bluck et al. 2014;2016). Here, AGN feedback is thought to be the major quenching mechanism,
and its strength is likely to depend on halo mass
(e.g. Croton et al. 2006; Henriques et al. 2016), qualitatively consistent with the observation results.

In addition to environmental effects that are confined within halos, there are also
observational indications that the environmental effects may operate on
scales beyond their boundaries. For example, at a fixed halo mass, the clustering of galaxy groups is found to
depend on the star formation rate and color of the central galaxies
(Yang, Mo, \& van den Bosch 2006; Lacerna, Padilla, \& Stasyszyn 2014). Similarly,
Kauffmann et al. (2013) found that the star formation rates of central galaxies
are correlated with that of their neighbours on scales up to several Mpcs,
far beyond their halo virial radii (see also Berti et al. 2017, and references therein).
These results suggest that large scale environments
may also affect the star formation of galaxies embedded in them.
However, it is unclear whether this is due to a causal connection between
star formation and large-scale environments, or is produced by a correlation induced by
some intermediate connections. For example, such large-scale effects
may be produced by the dependence of star formation on halo assembly history
(see e.g. Hearin, Watson, \& van den Bosch 2015; Lim et al. 2016;
Tinker et al. 2016; Zentner et al. 2016), combined with halo assembly
bias that links halo formation with large-scale
structure (e.g. Gao, Springel, \& White 2005; Wechsler et al. 2006; Jing, Suto \& Mo 2007),
or may be produced by the preheating of the intergalactic gas owing to the
formation of large scale structure (e.g. Mo et al. 2005; Kauffmann et al. 2013).

Using galaxy number density as environmental parameter, Baldry et al. (2006) found that the
quenched fraction depends both on galaxy stellar mass and environmental density,
and the dependence can be well described by a simple functional form.
Peng et al. (2010) studied the environmental quenching efficiency, which is defined
as the probability for a galaxy to be quenched in high-density regions
relative to that in low-density regions, where environmental effects are
expected to be weak. Remarkably, the efficiency defined in this way is found
to be almost independent of stellar mass. Subsequently, Peng et al. (2012) suggested that
the independence may be explained if environmental quenching is assumed to be important only for satellite
galaxies, and if both the quenching efficiency of satellite galaxies
and the satellite fraction  are independent of galaxy mass.
However, these assumptions are not supported by the results obtained
for centrals from galaxy groups, which clearly show that environmental quenching
of central galaxies is also important (e.g. Weinmann et al. 2006; Woo et al. 2013; 2015),
and by the results for satellites, which show that both the
quenching efficiency (Knobel et al. 2015) and satellite fraction
(Mandelbaum et al. 2006; Cooray 2006; Tinker et al. 2007;
van den Bosch et al. 2007) depend on galaxy mass.

The difference between centrals and satellites also attracted particular attention
(e.g. van den Bosch et al. 2008; Skibba 2009; Wetzel et al. 2012;2013; Peng et al. 2012; Hirschmann et al. 2014;
Knobel et al. 2015; Spindler \& Wake 2017). It has been found that the quenched fraction of the
central population is lower than that of satellites of the same mass. This difference has been used
to quantify the efficiency of various satellite-specific quenching processes,
such as strangulation, tidal stripping and ram-pressure stripping (van den Bosch et al. 2008).
However, Knobel et al. (2015) found that centrals and satellites of the same mass respond
to their environments in a similar way, as long as centrals have massive satellites. Moreover,
Hirschmann et al. (2014) studied the failures of current galaxy formation models
in matching observational data and suggested that centrals and satellites
should be treated not as differently in their response to environments
as previously assumed.

Clearly, more investigations are required in order to understand these contradictory
results in the literature. It is essential to identify and characterize the contribution to
the quenching of star formation of the relevant parameters, such as
galaxy stellar mass, halo properties and large-scale density field. In particular,
it is important to see whether the independence of environmental quenching efficiency
on galaxy mass can be reproduced if only halo masses and halo assembly
bias are taken into account, and what roles centrals and satellites
play in establishing the galaxy-mass independence of the quenching efficiency
found earlier, and whether the star formation quenching in centrals and satellites is dominated by different processes.

In this paper, the fourth of a series, we use the environmental information
provided by the ELUCID project and galaxy groups selected from the Sloan
Digital Sky Survey (SDSS; York et al. 2000) to investigate the quenching of
galaxies in different environments. The ELUCID project (Wang et al. 2014; 2016;
Tweed et al. 2017) aims to reconstruct the initial conditions responsible
for the formation of the structures in the observed low-redshift universe, and to
recover the mass distribution in the local universe by constrained simulations.
The constrained simulations give the full information about the dynamical
state and formation history of the large scale structure within which the observed
galaxies reside. This provides an unique opportunity to systematically investigate
the quenched population of galaxies of different masses in different environments.

Our paper is organized as follows. In Section \ref{sec_gs}, we describe the
galaxy sample, group catalog and environmental quantities used for our analysis.
Section \ref{sec_TQP} shows how the quenched galaxy population depends on
galaxy mass, halo mass and environmental density for the total population,
as well as separately for the central and satellite populations.
In Section \ref{sec_deqe}, we investigate the galaxy stellar mass independence
of the environmental quenching efficiency using the matter density field as
inferred from our ELUCID simulation, with a special focus on the role of central
and satellite galaxies. In Section \ref{sec_heqe}, we investigate the quenching
efficiencies using halo mass as environmental parameter. In Section \ref{sec_driver}
we discuss the implications of our results by comparing the data to three simple
models that incorporate the dependence on galaxy stellar mass, halo mass,
environmental density, and halo assembly history. In Section \ref{sec_cen} we discuss
whether central galaxies are special in their quenching properties
in comparison with satellites. Finally, we summarize our  results and discuss their implications in Section~\ref{sec_summary}.

\section{Galaxy Sample and Environmental Quantities}
\label{sec_gs}

\begin{figure*}
\centering
\includegraphics[width=0.8\textwidth]{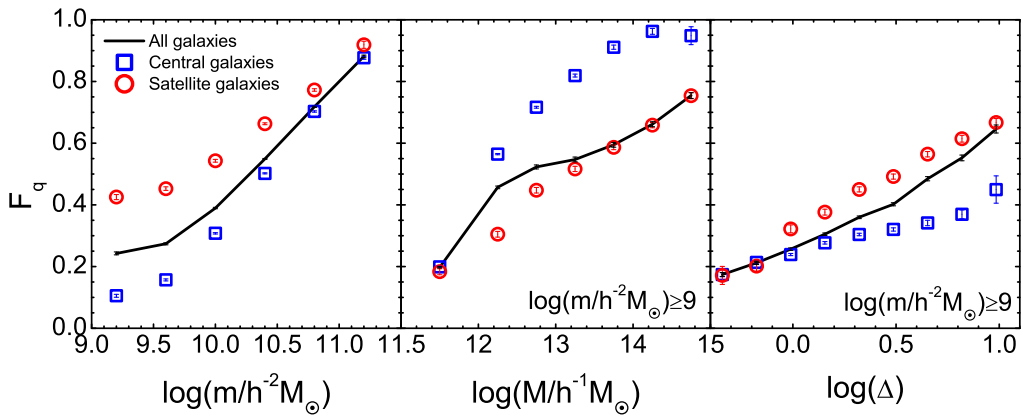}
\caption{Quenched fraction of galaxies as a
function of galaxy stellar mass (left panel),
halo mass (middle panel), and environmental density (right panel),
obtained from samples of the total (lines), central (squares) and satellite (circles)
populations. The middle and right panels
show the results only for galaxies with $\log(m/\msunt)\geq9.0$.}
\label{fig_qf}
\end{figure*}

\subsection{The galaxy sample}
\label{ssec_GGC}

The galaxy sample used here is extracted from the New York University Value-added
Galaxy Catalog (NYU-VAGC, Blanton et al. 2005) of the SDSS DR7 (Abazajian et al. 2009).
We select all galaxies in the main galaxy sample, with $r$-band apparent
magnitudes $\leq17.72$, and with redshift completeness $C\geq0.7$, and within the
reconstruction region of ELUCID simulation (see below and Wang et al. 2016 for the details).
The first two selection criteria ensure that most of the selected galaxies are contained
in the Yang et al. (2007) group catalog (with extension to DR7), and the third ensures
that we have reliable estimates of our galaxies' environmental densities.
Among these galaxies, 1,707 are members of groups that have only a fraction
of $0<f_{\rm edge}\leq 0.6$ of their virial volumes contained within the survey
boundary. These galaxies are removed from our sample (see Yang et al. 2007).
A total of 4,233 galaxies that do not have star formation rate estimates are
also discarded. Our final sample contains 317,791 galaxies.

Stellar masses for these galaxies, indicated by $m$ (with unit $\msunt$), are computed using the relations
between stellar mass-to-light ratio and $(g - r)$ color as given in
Bell et al. (2003), adopting a Kroupa (2001) initial mass function (IMF).
We refer to Yang et al. (2007) for details. Star formation rates (SFR) for these
galaxies are taken from the the MPA-JHU DR7 release
website\footnote{\tt http://www.mpa-garching.mpg.de/SDSS/DR7/}, which are estimated by using
an updated version of the method presented in Brinchmann et al. (2004) and calibrated
to the Kroupa IMF. For given galaxy stellar mass, the distribution of SFR is known to be bimodal (e.g. Brinchmann et al. 2004), with a high SFR mode corresponding to the star forming population and a low SFR mode corresponding to a quenched population.
In this paper, we adopt the division
line proposed by Woo et al. (2013) to separate the two populations:
\begin{equation}\label{eq_div}
\log{\rm SFR} =0.64\log{m}-1.28\log{h}-7.22\,,
\end{equation}
where the reduced Hubble constant $h$ (Hubble constant in units of
$100\,{\rm km\,s^{-1}Mpc^{-1}}$) is used to transfer the unit of the stellar
mass from $\msunt$ used here to $\Msun$ used in Woo et al. (2013).

In our analyses, each galaxy is assigned a weight $w=1/(V_{\rm max}C)$ to take into
account the Malmquist bias and redshift (spectroscopic) incompleteness, with the latter
taken from the NYU-VAGC. Since the geometry of our reconstruction region is not regular,
we calculate $V_{\rm max}$ in the following way. For each galaxy, we first obtain its
Petrosian photometry in $ugriz$ bands and its redshift. We then use these data as input
to the $K$-correction utilities (v4$\_$2) of Blanton \& Roweis (2007) to estimate
$z_{\rm min}$ and $z_{\rm max}$, the minimum and maximum redshifts, between which the
galaxy can be observed with the $r$-band limit of 17.72 mag. Finally, we measure $V_{\rm max}$ as the
volume of the reconstruction region that is between $z_{\rm min}$ and $z_{\rm max}$.

\subsection{Host halos of galaxies}

Galaxy groups and clusters (hereafter referred to together as galaxy groups),
when properly selected from a galaxy sample, can be used to represent
the host dark halos of galaxies. In this paper we make use of the galaxy groups
identified by Yang et al. (2012) from the SDSS DR7 to represent halos in which
galaxies reside. This group catalog
was constructed with the halo-based group finder developed by Yang
et al. (2005), which assigns new galaxies into groups based on
the size (virial radius) and velocity dispersion of the host dark
halo represented by the current members assigned to a tentative group.
Iterations are performed until the identification of member galaxies
as well as the estimation of halo mass converge.
The halo masses in the catalog are estimated via the ranking of two mass proxies: the total
luminosity or total stellar mass of all members brighter than $M_r=-19.5+5\log(h)$ in the
$r$-band. For our analysis, we adopt the halo masses, $M$ (with unit $\msun$), estimated using the
total stellar mass. Following common  practice, we define the central galaxy of a group to be the most
massive member, and all other members are referred to as satellites.
The reconstruction region is restricted to the redshift range $0.01<z<0.12$,
where groups with $\log(M)\gtsima 12$ are complete.

\subsection{Environmental density}
\label{ssec_Edensity}

In Wang et al. (2016; hereafter paper III), we presented a series of methods to
reconstruct the initial density field that is responsible for the galaxy distribution
in the local Universe, and we used a high-resolution N-body simulation to evolve the
initial conditions to the present-day. These simulation results can be used to
obtain reliable estimates for the environmental densities within which the observed
SDSS galaxies reside.

The reconstruction is restricted to the Northern Galactic Cap of the SDSS DR7 region
and to the redshift range $0.01\leq z\leq0.12$. In order to avoid problems near the
survey boundary, where the reconstruction is less reliable, we exclude
galaxies and groups whose distances to the SDSS survey boundary are smaller than 5$\mpc$.
For each galaxy, we correct for its redshift-space distortions (see Paper III, Wang et al.
2012 and Shi et al. 2016, for details) and estimate its real-space location.
The environmental density for a galaxy is determined by computing the matter
density, smoothed with a Gaussian kernel of 4$\mpc$, at its real-space location in
our constrained simulation. Tests based on mock galaxy catalogs demonstrate that the
uncertainties in this density estimate are typically $0.10$ dex. In what follows,
we quantify the matter density for each galaxy using the quantity
\begin{equation}
\Delta =\rho/{\bar\rho}\,,
\end{equation}
where $\rho$ is the smoothed mass density at the location of the galaxy,
and ${\bar\rho}$ is the mean density of the universe.

In the literature, one environmental indicator commonly used is
the local number density (or over-density) of galaxies. Our tests show that the galaxy densities are positively correlated with the mass
densities, although the scatter is rather large (see also Baldry et al. 2006). Because of the redshift distortion effect and the complicated correlation with underlying mass density, galaxy density hampers a meaningful interpretation of its correlation between galaxy
properties (see Weinberg et al. 2006 and Woo et al. 2013). Therefore, we consider the matter densities, $\Delta$, used here to be a superior, and more physical, quantity to characterize the large-scale environment of a galaxy.

\section{The quenched populations}
\label{sec_TQP}

\begin{figure*}
\centering
\includegraphics[width=0.9\textwidth]{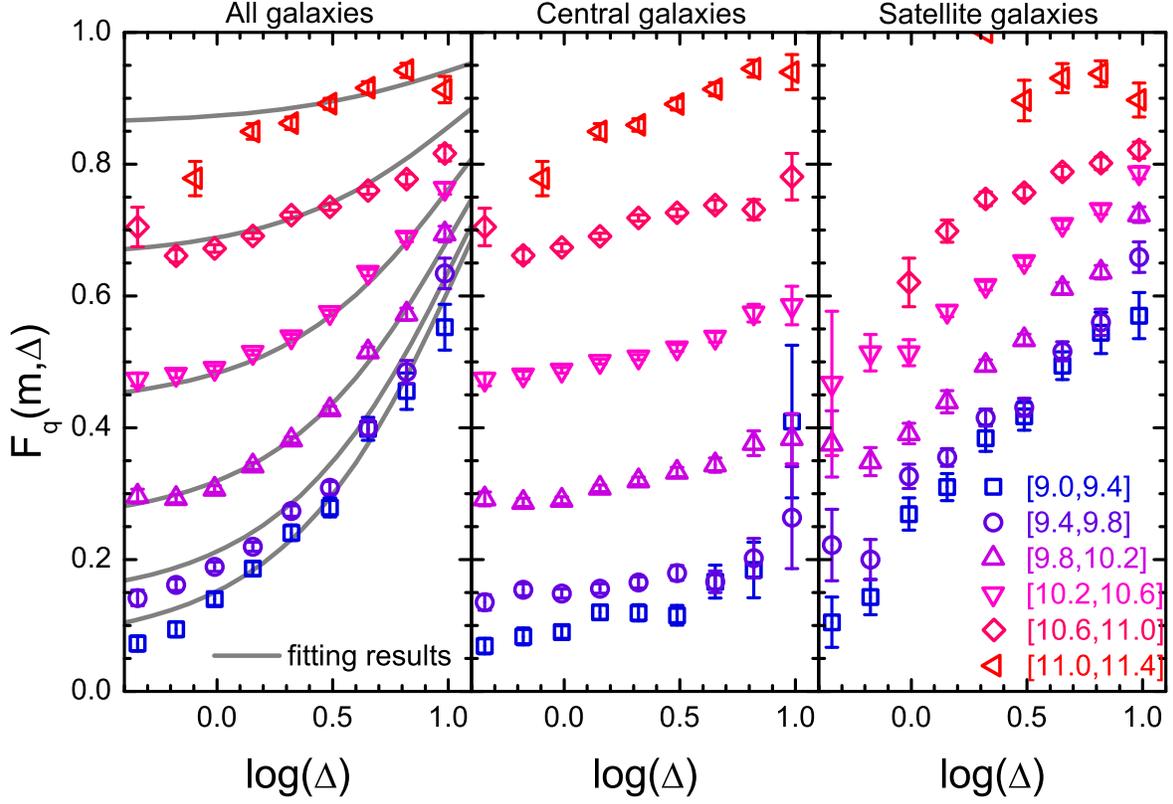}
\caption{Quenched fractions of galaxies in given stellar mass bins as
functions of the environmental density, obtained from samples of the
total (left panel), central (middle panel) and satellite (right panel)
populations. The curves in the left panel are model fits to the data
(see text).}
\label{fig_qfd4}
\end{figure*}

In this section we examine how quenching of star formation
depends on the intrinsic and environmental properties of galaxies.
For any subset of galaxies with a given set of properties $\{g\}$,
we calculate the average quenched fraction as
\begin{equation}\label{eq_fq3}
F_{\rm q}(\{g\})=\frac{\sum_{i\in \{g\}}q_iw_i}{\sum_{i\in\{g\}}w_i}\,,
\end{equation}
where $w_i$ is the weight for a galaxy $i$,
and $q_i$ is set to $1$ for a quenched galaxy, otherwise zero.  In our following analysis, we will
consider the total population, and the central and
satellite populations separately. The quenched fractions of the sub-populations
are denoted by $F_{\rm q, c}$ (centrals) and $F_{\rm q, s}$ (satellites), respectively.

\begin{figure*}
\centering
\includegraphics[width=0.8\textwidth]{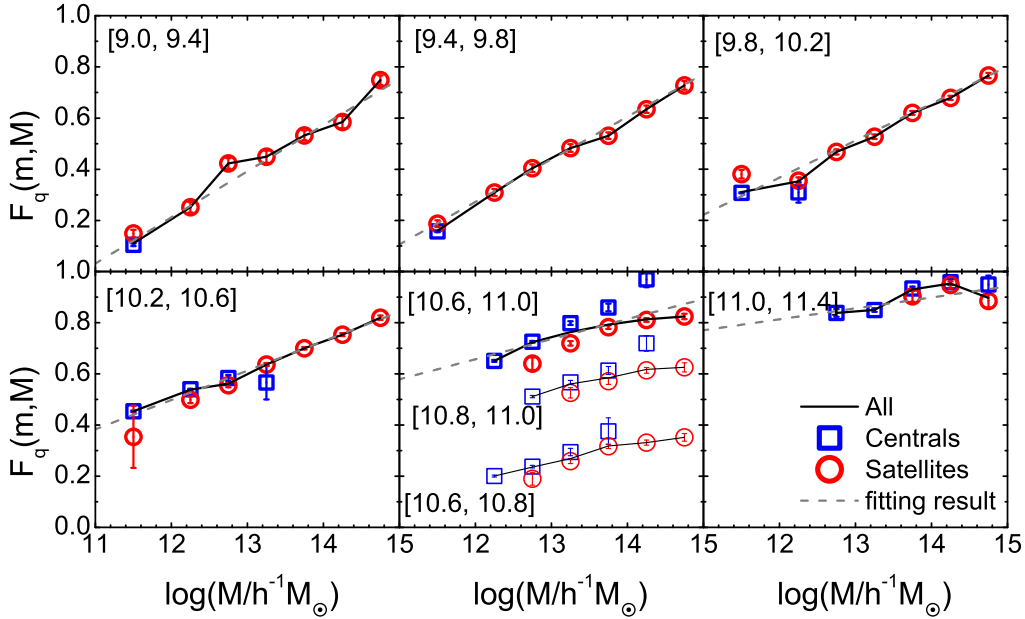}
\caption{Quenched fraction as a function of halo mass. Galaxies are divided into
six mass bins as shown in the panels. The squares and circles show the
results for centrals and satellites, respectively. The solid lines show the results for all
galaxies in the corresponding mass bins. The gray dashed lines show the fitting of
the total population with Equation (\ref{eq_fq4}). For galaxies with $10.6\leq\log(m)\leq11$,
we split them into two sub-samples with narrower mass ranges as indicated in the panel.
For clarity, the two sub-samples are shifted down by 0.25 for high mass bin and 0.45 for low mass bin,
respectively.}
\label{fig_qfmh}
\end{figure*}

\subsection{Marginalized dependence on stellar mass, halo mass and density}

Figure \ref{fig_qf} shows, for the whole population and for centrals and satellites separately, 
the quenched fraction of galaxies as a function of galaxy stellar mass, halo mass and environmental mass density.
The quenched fractions as a function of one of the three parameters are obtained by marginalizing
over the other two parameters. Note that we only consider galaxies with $\log(m)\geq9.0$
and results are shown only for these galaxies in the left two panels.
In this and the following figures, error bars are all evaluated from 1,000 bootstrap re-samplings and only
data points that contain at least 10 galaxies are shown.

It is clear that the quenched fraction increases with increasing stellar mass, halo mass and
environmental density, consistent with previous studies (e.g. Brinchmann et al. 2004; Weinmann et al. 2006;
Baldry et al. 2006).  These correlations are often
considered as observational evidence for quenching processes, such as mass quenching,
halo quenching and environmental quenching. We then examine centrals and satellites separately. At given stellar mass or density, satellites tend to
be more often quenched than centrals. However, the trend is reversed if halo mass is used instead as the control parameter.
It is well known that more massive galaxies tend to live in more massive halos and
more massive halos tend to reside in higher density regions, and so it is not obvious
whether the density dependence reflects a causal connection or is induced by
the correlation between halo mass and density. In order to disentangle the
different effects, we need to examine the joint distribution of the quenched
population with respect to the parameters in question (see below).

\subsection{Dependence on stellar mass and environmental density}\label{ssec_qfd}

We first disentangle the dependencies of the quenched fraction on
stellar mass and environmental density.
The left panel of Figure \ref{fig_qfd4} shows the quenched fraction
of the total population, $F_{\rm q}(m,\Delta)$, as a function of
the environmental density for galaxies in various stellar mass bins.
Consistent with the marginalized result, there is
significant dependence on $\Delta$,
even for galaxies in a given narrow $m$ bin. The dependence is
seen to be stronger at low $m$ and weaker at the lower $\Delta$ end. At a given $\Delta$,
the quenched fraction increases with $m$, and the increase is
more significant for galaxies of lower masses.

We then investigate the quenched fractions separately for the central and satellite populations and present the results in
the middle and right panels of Figure \ref{fig_qfd4}.
For central galaxies, the $\Delta$-dependence is rather weak for all stellar
mass bins in comparison to the result shown in the right panel of Figure \ref{fig_qf}.
This suggests that the marginalized $\Delta$-dependence for centrals is primarily
due to the fact that more massive centrals tend to reside in higher density region.
In contrast, for satellites, the $\Delta$-dependence is
strong in most of the stellar mass bins. In particular, the
marginalized $\Delta$-dependence is very similar to that for the lowest
stellar mass galaxies, indicating that the marginalized result
for satellites is dominated by low-mass galaxies over the whole density range.
At given $\Delta$, the $m$-dependence is
significant for both centrals and satellites. The $m$-dependence for centrals
is similar to that for satellites at low $\Delta$, but is stronger at high $\Delta$,
suggesting that the difference shown in the marginalized $m$-dependence
between the two populations is mainly caused by the difference in the
high density region. These results are in qualitative agreement with the
results previously obtained by Peng et al. (2010; 2012)
and Knobel et al. (2015).

\begin{figure*}
\centering
\includegraphics[width=0.8\textwidth]{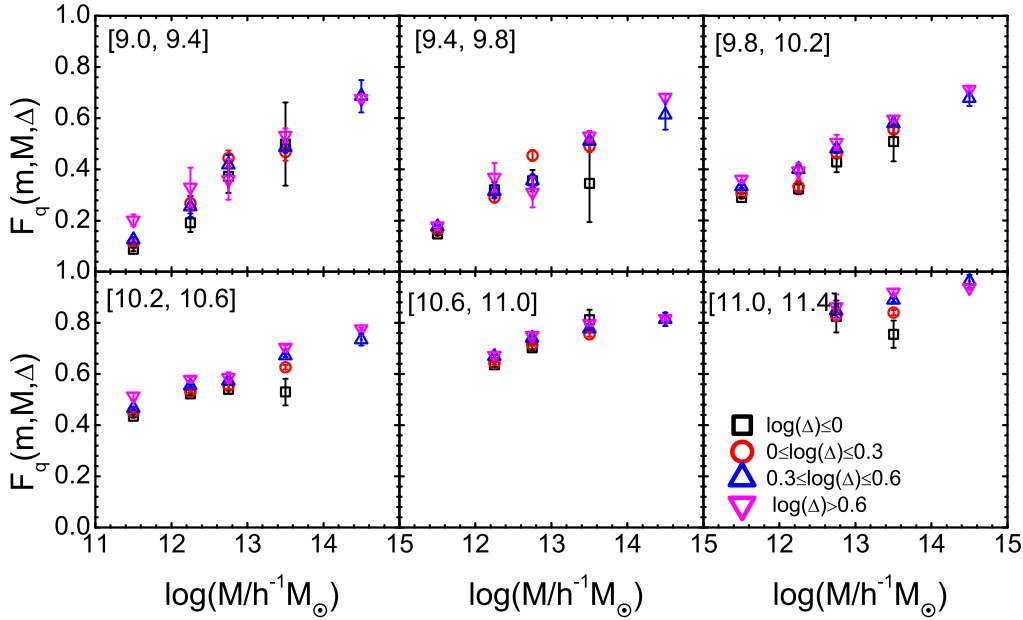}
\caption{Quenched fraction as a function of halo mass for galaxies with different $\Delta$ and
$m$ as indicated in the panels.}
\label{fig_qfmhrho}
\end{figure*}

\subsection{Dependence on stellar mass and halo mass}
\label{ssec_qfh}

Figure \ref{fig_qfmh} shows the quenched fraction as a function of halo mass  for galaxies in different stellar
mass bins. At a given $m$, the quenched fraction increases with $M$, and the increase is steeper for lower mass galaxies.
These trends are consistent with those obtained before
(e.g. Weinmann et al. 2006;  Wetzel et al. 2012). Most intriguingly, the
behaviors of central and satellite galaxies are
almost indistinguishable, although it is important to stress that
the halo mass range covered by central galaxies is rather limited,
in particular for galaxies with low masses. This suggests that the difference
between centrals and satellites found in the marginalized dependencies on
$m$ and $M$ are mainly caused by the different $m$ and $M$
ranges covered by the two populations. For example, satellites usually
reside in more massive halos than centrals of the same stellar mass,
this combined with the $M$-dependence of the quenched fraction can explain
why the quenched fraction for satellites is higher than that for
centrals of the same stellar mass (left panel of Figure \ref{fig_qf}).
Similarly, since by construction centrals are more massive than satellites
in halos of a given mass, centrals are expected to be more often quenched than
satellites because of the $m$-dependence (middle panel of
Figure \ref{fig_qf}).

For the stellar mass bin $10.6\leq\log(m)\leq11$, the quenched fraction of centrals
appears to be somewhat higher than for satellites in the same halo
mass bin. However, upon closer inspection, we find that this is
mainly an artifact of the finite bin-sizes used, combined with the
fact that at given halo mass, the satellite galaxies are strongly biased
towards $\log(m)\sim10.6$, while centrals are biased towards
$\log(m)\sim11$. To demonstrate this, we split the galaxies
in this mass bin into two sub-samples at $\log(m)=10.8$.
The results for these two sub-samples are shown in the bottom left panel
of Figure \ref{fig_qfmh}. As one can see, this significantly reduces the
differences between the centrals and satellites, and we therefore conclude
that there is no significant difference in the halo mass dependence
of the quenched fraction between centrals and satellites.

We note again that numerous previous studies (e.g., van den Bosch et al. 2008; Weinmann et al. 2009;
Pasquali et al. 2010; Wetzel et al. 2012; Peng et al. 2012; Knobel et al. 2013;
Bluck et al. 2014; Fossati et al. 2017) have shown that, at a given stellar mass,
satellites are more often quenched than centrals and suggested that some satellite-specific environmental processes have played important roles in quenching satellite galaxies.
However, our results portray a different picture; centrals and satellites follow the same correlation  between quenched
fraction and host halo mass, suggesting that they experienced similar environment-dependent
quenching, and the only reason that satellite and centrals appear different at fixed stellar mass is
that they sample different ranges in host halo mass. Further investigations
on this are presented in Sections \ref{sec_driver} and  \ref{sec_cen}.

\subsection{Dependence on halo mass and environmental density}
\label{ssec_QFHMED}

It is known that more massive halos tend to locate in higher density regions,
an effect usually referred to as halo bias (e.g. Mo \& White 1996). Therefore
the density dependence shown in Figure~\ref{fig_qfd4} and the halo mass dependence
shown in Figure~\ref{fig_qfmh} may be connected. To examine this, we split galaxies
further into four sub-samples based on $\Delta$ and
calculate the quenched fraction as a function of $M$. The results are shown in
Figure \ref{fig_qfmhrho}. Overall, the $M$-dependence of the
quenched fraction is similar for different sub-samples of $\Delta$,
suggesting that halo mass may be the dominating factor in determining the
quenched fraction.

However, for a given $M$, there is a weak but systematic
trend  of increasing quenched fraction with increasing $\Delta$.
This suggests that factors other than halo mass affect galaxy quenching, and
is broadly consistent with some previous findings that galaxy groups of a
given halo mass have different clustering properties depending on the color
or star formation of their member galaxies (e.g. Yang et al. 2006).
All of this is most likely related to assembly bias (Gao et al. 2005), the fact that halo bias
depends not only on halo mass, but also on other halo properties such
as assembly history, if the quenching processes depend on these additional halo properties. Unfortunately,
the observational sample is still too small, and the uncertainties in
the density dependence shown in Figure \ref{fig_qfmhrho} are still too large,
to draw any quantitative conclusions. We will revisit
this issue in Section \ref{sec_driver}.

\section{Density-based environmental quenching efficiencies}
\label{sec_deqe}

\begin{figure*}
\centering
\includegraphics[width=0.9\textwidth]{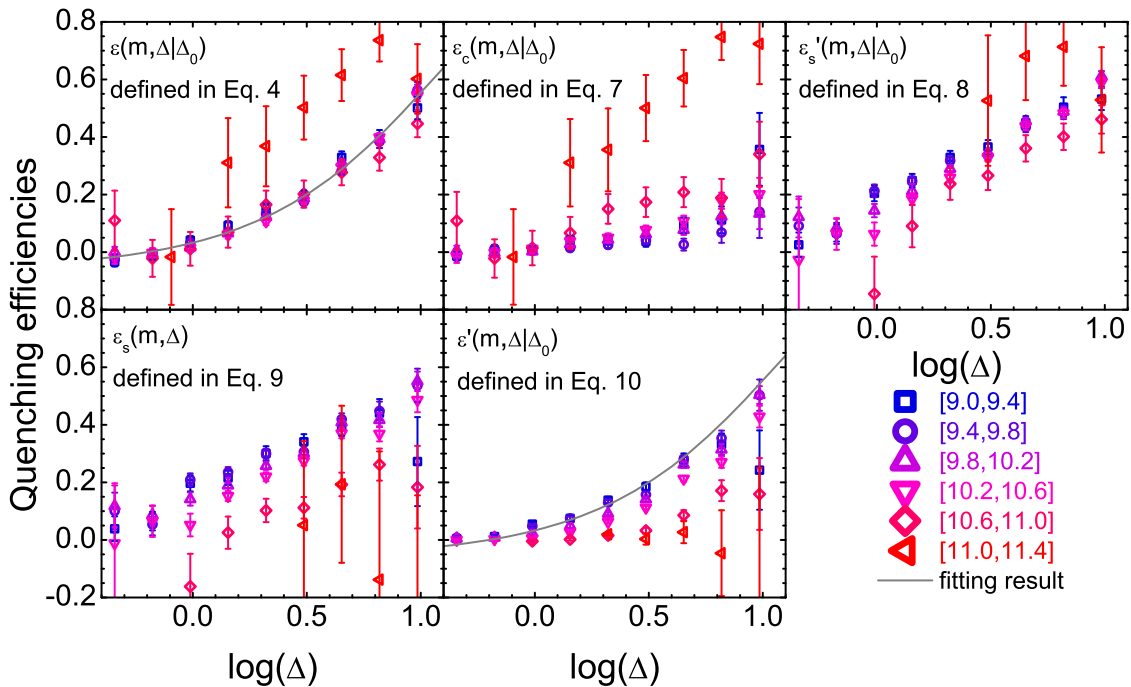}
\caption{Quenching efficiencies defined by the equations shown in each panel as a function of
environmental density $\Delta$ for galaxies of different masses as indicated in the bottom right panel.
The gray line shown in the top left and bottom middle panels is the fitting curve
[fitting formula Equation (\ref{eq_dee2})] of the data shown in the top left panel. See the text for the details.
Note that, for the most massive galaxies, two data points at $\log\Delta\sim0.3$ and 1.0 lie outside
of the figure boundary in the bottom left panel, and one data point at $\log\Delta\sim1.0$ lies outside the boundary in the bottom middle panel.}
\label{fig_effdv}
\end{figure*}

A useful parameter to quantify the efficiency of galaxy quenching
is the relative environmental quenching efficiency (hereafter `quenching
efficiency' for brevity)
\begin{equation}\label{eq_dee1}
\varepsilon(m, \Delta | \Delta_0) \equiv
{F_{\rm q}(m,\Delta)-F_{\rm q}(m,\Delta_0)\over 1-F_{\rm q}(m,\Delta_0)}\,.
\end{equation}
(Peng et al. 2010), which specifies the probability for a star-forming
galaxy of mass $m$ to be quenched when it transits from an environment
characterized by some zero-point matter density, $\Delta_0$, to a region
with $\Delta$. We choose $\Delta_0$
to correspond to the lowest-density environment probed by our data, as the environmental effects are expected
to be minimal in these void-like environments. As shown in Figure \ref{fig_qfd4},
the quenched fraction of galaxies, in most of the stellar mass bins, has the
lowest value at $\log\Delta\leq 0$, where the $\Delta$-dependence of the quenched
fraction is also weak. We thus choose galaxies with $\log\Delta\leq0$ to define
the zero point.

\subsection{A dearth of stellar mass dependence for the total population}
\label{ssec_dearth}

\begin{figure*}
\centering
\includegraphics[width=0.9\textwidth]{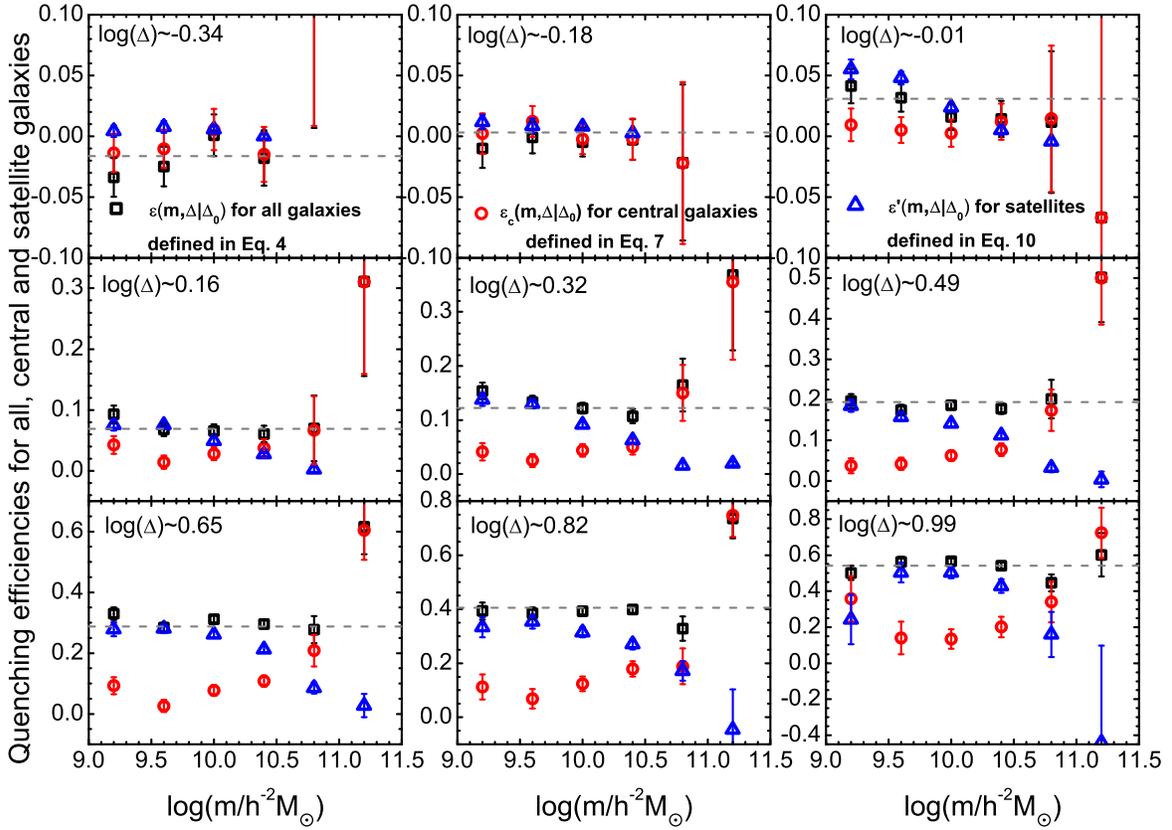}
\caption{The quenching efficiencies defined by Equation (\ref{eq_dee1}), (\ref{eq_dec}) and (\ref{eq_dee3}) as a function of
stellar mass for various environmental densities as shown in each panel. The horizontal dashed lines show the best fitting results of Equation (\ref{eq_dee2}). In the top-middle panel, two data points ($\varepsilon$ and $\varepsilon_{\rm c}$) at $\log m=11.2$ lies outside the figure boundary.}
\label{fig_effdv2}
\end{figure*}

The upper-left panel of Figure \ref{fig_effdv} shows the quenching efficiency
$\varepsilon(m, \Delta | \Delta_0)$ of the total population as a function of $\Delta$. As one can see,
the efficiency is close to zero at $\Delta\sim\Delta_0$, by definition,
and increases rapidly with $\Delta$. Remarkably,  the efficiency is almost
independent of $m$ over the range $9\leq\log(m)\leq11$.
To see this more clearly, we show $\varepsilon(m, \Delta | \Delta_0)$ as a
function of $m$ for various $\Delta$ bins in Figure \ref{fig_effdv2} as
black squares. The quenching efficiency is almost a constant in the
entire mass range,  except for the highest stellar mass bin.
This is in good agreement with Peng et al. (2010), although they used
galaxy number density instead of the more physical matter density used here.
The efficiency for the highest stellar mass bin is higher than for the other
stellar mass bins, but the uncertainties are large. Note that there
are only very few massive galaxies in low density regions with
$\Delta < 1$, which can produce a large statistical uncertainty in the
denominator of Eq.~(\ref{eq_dee1}). The discrepancy for the highest
$m$ bin is, therefore, not conclusive.
We will come back to this issue in Section \ref{sec_driver}.

Baldry et al. (2006) found that the quenched fraction can be fitted by a very simple equation:
\begin{equation}\label{eq_fq1}
F_{\rm q}(m,\Delta)=1-\exp\left[-\left(\frac{\Delta}{\Delta_{\star}}\right)^b\right]
\exp\left[-\left(\frac{m}{m_{\star}}\right)^d\right]\,.
\end{equation}
Inserting it into Equation (\ref{eq_dee1}), we obtain
\begin{equation}\label{eq_dee2}
\varepsilon(m, \Delta | \Delta_0)=1-\exp\left[-\left(\frac{\Delta}{\Delta_{\star}}\right)^b
+\left(\frac{\Delta_0}{\Delta_{\star}}\right)^b\right]\,.
\end{equation}
We use this equation to fit the $\varepsilon$ data points, with $\log\Delta_0=-0.2$,
which is the median value for galaxies with $\log\Delta_0 \leq 0$.
The best fit gives $\Delta_{\star}=11.5\pm0.26$ and $b=0.975\pm0.035$ (see Figure \ref{fig_effdv} for the fitting curve).
We then use Equation (\ref{eq_fq1}) to fit the quenched fraction for the total population with $\Delta_{\star}=11.5$ and $b=0.975$.
The best fit values for the other two parameters are $m_{\star}=5.29\pm0.04\times10^{10}\msunt$ and $d=0.749\pm0.006$, and the results
are shown in the left panel of Figure \ref{fig_qfd4}.
These fitting results are presented here as a convenient way
to represent the data.

\subsection{Central versus satellite populations}\label{ssec_ecs}

For central galaxies, we define a quenching efficiency similar to Equation (\ref{eq_dee1}):
\begin{equation}\label{eq_dec}
\varepsilon_{\rm c}(m, \Delta | \Delta_0) =
{F_{\rm q,c}(m,\Delta)-F_{\rm q,c}(m,\Delta_0)\over 1-F_{\rm q,c}(m,\Delta_0)}\,,
\end{equation}
where $F_{\rm q,c}(m,\Delta)$ is the quenched fraction of central galaxies. Here again we use galaxies at
$\log\Delta_0\leq 0$ to calculate $F_{\rm q,c}(m,\Delta_0)$. So defined, this
efficiency characterizes the probability for a star forming central to quench
if it were to move to a higher density environment,
$\Delta$, while remaining a central. As is evident from the upper-middle panel of Figure \ref{fig_effdv},
$\varepsilon_{\rm c}$ has a much weaker dependence on $\Delta$ than the
corresponding quenching efficiency of the total population. Nevertheless, even for centrals there is a significant
tendency for $\varepsilon_{\rm c}$ to increase with $\Delta$. Moreover, there
is also a trend for $\varepsilon_{\rm c}$ to increase with stellar mass (see also Figure \ref{fig_effdv2}), in particular at high $\Delta$. As for the total population, the most massive bin has very few galaxies with
$\log\Delta\leq0$, so that the corresponding $\varepsilon_{\rm c}$ carries
large uncertainties.

Similarly, we can define a quenching efficiency for satellite galaxies. Unfortunately, the total
number of satellites at $\log\Delta\leq 0$ is small, and so the derived efficiency will
have large uncertainties. However, $F_{\rm q,s}(m,\Delta_0)$ is close to
$F_{\rm q,c}(m,\Delta_0)$ for most of stellar mass bins where both can be measured
reliably (Figure \ref{fig_qfd4}). We thus define an alternative efficiency for satellites as
\begin{equation}\label{eq_des1}
\varepsilon'_{\rm s}(m, \Delta | \Delta_0) =
{F_{\rm q,s}(m,\Delta)-F_{\rm q,c}(m,\Delta_0)\over 1-F_{\rm q,c}(m,\Delta_0)}\,.
\end{equation}
The results are shown in the upper-right panel of Figure \ref{fig_effdv}.
As one can see, $\varepsilon'_{\rm s}(m, \Delta | \Delta_0)$
increases rapidly with $\Delta$, but its dependence on $m$ is weak.

Finally we consider a satellite-specific quenching efficiency,
which is defined as
\begin{equation}\label{eq_des2}
\varepsilon_{\rm s}(m, \Delta) = \frac{F_{\rm q,s}(m,\Delta)-F_{\rm q,c}(m,\Delta)}
{1-F_{\rm q,c}(m,\Delta)}\,,
\end{equation}
(see van den Bosch et al. 2008). Here, the central galaxies are
used as the control sample (zero point) in $(m, \Delta)$ space,
against which the quenching of satellites is measured. The lower-left panel of
Figure \ref{fig_effdv} shows $\varepsilon_{\rm s}$ as a function
of $\Delta$ for satellite galaxies of different masses. As one
can see, the quenching efficiency increases quite rapidly with
$\Delta$ in most stellar mass bins. For a given $\Delta$,  $\varepsilon_{\rm s}$
decreases with $m$, and the decrease is larger for more
massive galaxies.

As pointed out by Wetzel et al. (2013), $\varepsilon_{\rm s}$ defined in this way
actually measures a combined effect of the satellite-specific
quenching processes and the evolution of central galaxies.
So it is not straightforward to use it to interpret the satellite quenching efficiency.
However, it does not change our conclusion in the subsequent subsection that there exists
an unexpected connection between centrals and satellites.

\subsection{A conspiracy between centrals and satellites?}\label{ssec_csp}

\begin{figure}
\centering
\includegraphics[width=0.5\textwidth]{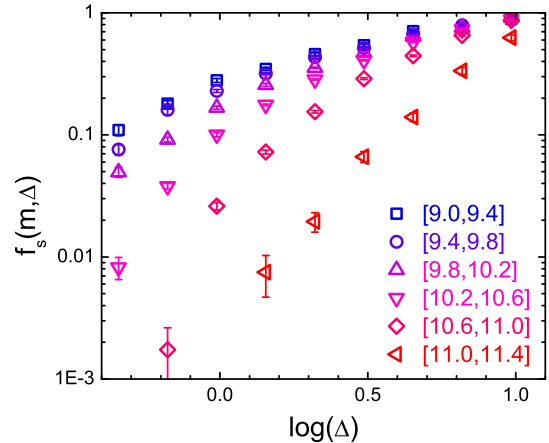}
\caption{The satellite fraction as a function of $\Delta$ at various galaxy masses. Note
that the two data points at low $\Delta$ for the two most massive galaxy bins are zero,
and thus not shown in the figure.}
\label{fig_fsd}
\end{figure}

Peng et al. (2010) argued that the stellar mass independence of
$\varepsilon(m, \Delta | \Delta_0)$ can be fully understood in terms
of the quenching of satellite galaxies. Assuming that the
environmental effect on centrals is negligible, they wrote the
environmental quenching efficiency as,
\begin{equation}\label{eq_dee3}
\varepsilon'(m, \Delta | \Delta_0)=f_{\rm s}(m, \Delta)
\varepsilon_{\rm s}(m, \Delta)\,,
\end{equation}
where, $f_{\rm s}(m, \Delta)$ is the satellite fraction, and $\varepsilon_{\rm s}(m, \Delta)$
is the satellite-specific quenching efficiency of Equation.~(\ref{eq_des2}).
Figure~\ref{fig_fsd} shows the satellite fraction as a function of
$\Delta$ for galaxies in different $m$ bins.
In order to explain the $m$-independence of the quenching
efficiency, Peng et al. (2010; 2012) suggested that both $f_{\rm s}(m, \Delta)$
and $\varepsilon_{\rm s}(m, \Delta)$ are independent of $m$.
This is clearly inconsistent with our data shown in the lower-left panel of Figure \ref{fig_effdv} and Figure~\ref{fig_fsd}.

Since the hypothesis made by Peng et al. has far-reaching implications (i.e., the the environment dependence
of the quenching efficiency is entirely due to the quenching of satellites),
it is important to address this discrepancy in some detail. In fact, there are
a number of factors that play a role. First of all, Peng et al. used the
over-density of galaxies as their environment indicator, rather than the more
physical matter density used here. In addition, they used colors to
split their population into star-forming and quenched, whereas we use actual
star formation rates. Since dust extinction can make a star forming galaxy
appear red, and thus `quenched' based on color, using actual star formation rates
yields more accurate estimates of the true quenched fraction. Secondly,
Peng et al., assumed $F_{\rm q,c}(m,\Delta)$ to be independent of $\Delta$,
even though this is not supported by their own data. Since centrals on average reside
in lower $\Delta$ environments than satellites of the same stellar mass
(e.g. Knobel et al. 2015), the average of $F_{\rm q,c}$ over $\Delta$ is
biased towards low-density regions. As a consequence, adopting the average
of $F_{\rm q,c}$ can lead to an overestimation of $\varepsilon_{\rm s}$, in particularly at
the high-$\Delta$ end. For low-mass galaxies, this bias is negligible,
because $F_{\rm q,c}$ is on average much smaller than both $F_{\rm q,s}$
and unity. This is consistent with the weak $m$-dependence of $\varepsilon_{\rm s}$
we find for these galaxies. For massive galaxies, on the other hand,
the value of $F_{\rm q,c}$ is higher and the denominator in
Equation~(\ref{eq_des2}) smaller. This results in a significant bias
that weakens the $m$-dependence of $\varepsilon_{\rm s}$ at the high mass end.
Indeed, when taking into account the dependence of $F_{\rm q,c}$ on galaxy number density,
Knobel et al. (2015) also found that $\varepsilon_{\rm s}$ decreases with
$m$ at a given galaxy number density (see their figure 3).
Finally, as is evident from Figure~\ref{fig_fsd}, the satellite fraction depends strongly on $m$. In fact, it is well known
that the satellite fraction increases with decreasing stellar mass, which
has been demonstrated using galaxy group catalogs (e.g., van den Bosch et al. 2008),
subhalo abundance matching (e.g., Wetzel et al. 2013), galaxy-galaxy lensing
(e.g., Mandelbaum et al. 2006), galaxy clustering (e.g., Cooray 2006; Tinker
et al. 2007; van den Bosch et al. 2007) and combinations thereof (e.g.,
Cacciato et al. 2013).

In the lower-middle panel of Figure \ref{fig_effdv}, we show $\varepsilon'$
as a function of $\Delta$. For low-mass galaxies with $\log(m)\leq 10.2$,
$\varepsilon'$ is almost independent of $m$, in agreement with
the hypothesis of Peng et al. (2010; 2012). However, for more massive galaxies,
$\varepsilon'$ clearly depends on $m$. Thus, the $m$-independence
of $\varepsilon$ of the total population shown in Section \ref{ssec_dearth} is not due
to the $m$-independence of $f_{\rm s}$ and $\varepsilon_{\rm s}$, and the quenching
efficiency of the total galaxy population cannot be explained by the quenching of satellites alone.

As shown in Figure \ref{fig_qfd4}, central galaxies do exhibit some weak but non-trivial
$\Delta$-dependence in all the mass bins considered, and at
$\log(m)\geq 10.6$ the dependence is actually comparable to or slightly stronger
than that for satellites. For centrals, the $\Delta$-dependence is stronger for
more massive galaxies, a trend opposite to that seen for satellites.
To understand the importance of quenching for centrals in the total population,
we express the quenching efficiency for the total population in terms of
$\varepsilon_{\rm c}$ and $\varepsilon'$.
At very low density, $\log\Delta \leq 0$, environmental quenching is very weak
and $f_{\rm s}(m, \Delta)$ is quite small. Thus, to good approximation
$F_{\rm q}(m,\Delta_0)\approx F_{\rm q,c}(m,\Delta_0)$, and we can
rewrite Equation\,(\ref{eq_dee1}) as
\begin{eqnarray}\label{eq_dee4}
&\varepsilon&(m, \Delta  | \Delta_0)
\simeq \frac{F_{\rm q}(m,\Delta)-F_{\rm
         q,c}(m,\Delta_0)}{1-F_{\rm q,c}(m,\Delta_0)}\nonumber\\
&=&\varepsilon_{\rm c}(m, \Delta | \Delta_0)+[1-\varepsilon_{\rm c}(m, \Delta | \Delta_0)]
   \varepsilon'(m, \Delta | \Delta_0)\,,
\end{eqnarray}
where the second equation is obtained by inserting Equations (\ref{eq_dec}) and (\ref{eq_des2})
 into the right-hand-side of the first line.
Note that this reduces to Equation (\ref{eq_dee3}) in the limit
$\varepsilon_{\rm c} \rightarrow 0$. Using
Equation (\ref{eq_dee4}) to compute the quenching efficiencies yields the
results that are in excellent agreement
with those shown in Figure \ref{fig_effdv} obtained using the original
definition [Equation.~(\ref{eq_dee1})]. Since the difference in all $m$ bins is less than 0.03, we do not show the results.
The good agreement justifies our
approximation that $F_{\rm q}(m,\Delta_0)\approx F_{\rm q,c}(m,\Delta_0)$.

Note once more that $\varepsilon(m, \Delta  | \Delta_0)$ reveals only
a very weak dependence on $m$, except for the most massive galaxies,
for which the statistics is extremely poor (see Section \ref{ssec_dearth}).
It seems to conflict with the fact that $\varepsilon$ is the combination of $\varepsilon_{\rm c}$ and $\varepsilon'$, which both strongly depend on $m$. To understand this apparent discrepancy, we show $\varepsilon$, $\varepsilon_{\rm c}$ and $\varepsilon'$ as a function of $m$
for various $\Delta$ bins in Figure \ref{fig_effdv2}. Apparently, the opposite trends in the $m$-dependence of
$\varepsilon_{\rm c}$ and $\varepsilon'$ counterbalance each other
so as to yield a $\varepsilon$ that depends only weakly on $m$.

Our results clearly demonstrate that the environmental effect on centrals has to be
taken into account in order to reproduce the $m$-independence of the quenching
efficiency seen for the total population. This is particularly important for massive
centrals. Peng et al. (2012) also found a significant environmental dependence for
central galaxies. However, they suspected that it is caused by the misidentification of satellites as centrals.
We indeed find some signals for such misidentification in our results. For example, some abnormal
behavior of the lowest mass galaxies in high-density bins can be seen in
$\varepsilon_{\rm s}$, as well in $\varepsilon_{\rm c}$.
More recently, Hirschmann et al. (2014) investigated the misidentification problem
using mock galaxy catalogs constructed from a semi-analytic model of
galaxy formation, and found that the contamination in centrals is less than 10\%
for most galaxy masses and environmental densities. The average contamination is
less than 6.5\%, roughly independent of $m$ (see also Lange et al. 2017). Since the
difference between
$F_{\rm q,c}$ and $F_{\rm q,s}$ rapidly decreases with increasing $m$
(Figure \ref{fig_qfd4}), the impact of the contamination is expected to decrease
with $m$. In contrast, the observed $\Delta$-dependence is more important for more
massive galaxies. Based on these results, we are confident that central-satellite
contamination does not significantly impact our conclusion that satellite
quenching alone cannot account for the  $m$-independence of the quenched efficiency,
at least for massive galaxies.

An interesting question is why the increase of $\varepsilon_{\rm c}$ with
increasing $m$ apparently compensates the decrease of $\varepsilon'$
with increasing $m$. It might reflect some deeper
connection between the quenching processes for centrals and satellites.
As shown above, the quenched fractions of centrals and satellites of the same
stellar mass correlate with the environment as characterized by halo mass in
the same way. This suggests that centrals and satellites may experience
similar quenching processes which are ultimately related to the host halo mass. In Section \ref{sec_driver} we will
construct simple models to investigate this issue.

\section{Halo-based quenching efficiencies}
\label{sec_heqe}

As discussed earlier, halos play a crucial role in shaping galaxy properties.
Many quenching processes are thought to correlate with halo mass and
stellar mass (see Mo et al. 2010). In order to examine this, we define
two new quenching efficiencies in the same vein as the environmental quenching
efficiency of Equation (\ref{eq_dee1}). These are the halo-based
environmental quenching efficiency, defined as
\begin{equation}\label{eq_hee1}
\varepsilon(m, M | M_0)
\equiv {F_{\rm q}(m,M)-F_{\rm q}(m,M_0)\over 1-F_{\rm q}(m,M_0)}\,,
\end{equation}
and the stellar mass quenching efficiency, defined as (see also Peng et al. 2010)
\begin{equation}\label{eq_hme}
\varepsilon_{\rm m}(m, M | m_0)
\equiv {F_{\rm q}(m,M)-F_{\rm q}(m_0,M)\over 1-F_{\rm q}(m_0,M)}\,.
\end{equation}
Here $M_0$ and $m_0$ are the halo mass and stellar mass `zero-points' against
which the dependencies on $M$ and $m$ are compared. Note that
$\varepsilon(m, M | M_0)$ and $\varepsilon_{\rm m}(m, M | m_0)$ characterize
the dependence on halo mass and stellar mass, respectively, of the combined effect
of all quenching processes, including `environmental processes' (such as
ram-pressure/tidal stripping and strangulation) and `internal processes' (such as
quenching induced by AGN and supernova feedback).

\subsection{Quenching efficiencies for the total population}\label{ssec_mhmg}

\begin{figure}
\centering
\includegraphics[width=0.5\textwidth]{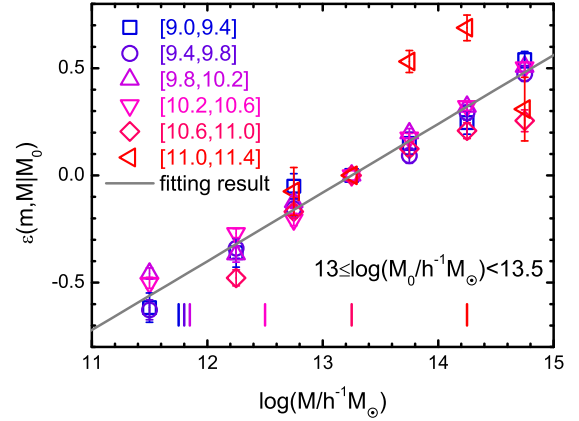}
\caption{The symbols with error bars show the environmental quenching
efficiency $\varepsilon(m, M | M_0)$ [see Equation (\ref{eq_dee1})] as a function of
halo mass $M$ for galaxies of different masses, as indicated in the panel.
The efficiency is calculated by using galaxies with
$13\leq\log{M_0/\Msun}< 13.5$ to estimate $F_{\rm q}(m,M_0)$. The gray line is the fitting
curve (Equation \ref{eq_emh}) of the data. The short vertical lines
mark the halo masses at which galaxies in a given stellar mass bin transit from
central dominated to satellite dominated.}
\label{fig_emh}
\end{figure}

Ideally, one would like to adopt the lowest halo mass as `zero-point' environment to
calculate the environmental efficiency. However, because no massive galaxy
($\log(m)>10.6$) resides in halos of $\log(M)<12$, choosing
the lowest halo mass bin is inappropriate. We therefore adopt $13\leq\log(M)<13.5$
to define the quenching zero-point, at which the estimates of the quenched fraction
are robust for all stellar mass bins (see Figure \ref{fig_qfmh}). The corresponding
efficiency, obtained from Equation (\ref{eq_hee1}), is shown as a function of halo
mass in Figure \ref{fig_emh}. Here again the environmental quenching efficiency is
almost independent of stellar mass, although it increases strongly with halo mass.
It is easy to see that, if the efficiency is independent of stellar mass, this
independence holds regardless of the value of $M_0$ used. This result clearly
shows that the environmental dependence of the quenched fraction can be well
separated from the dependence on stellar mass, independent of whether
the environmental parameter is the large scale matter density or the mass of the
host halo in which the galaxies reside.

Motivated by Equation (\ref{eq_fq1}) and that the dependence of $F_{\rm q}$ on halo mass
can be well described by a power law function (Figure \ref{fig_qfmh}), we propose to use a simple formula to describe the quenched fraction:
\begin{equation}\label{eq_fq4}
F_{\rm q}(m,M)=1-\exp\left[-\left(\frac{m}{m_{\star}}\right)^d\right](a\log M-ac)\,.
\end{equation}
It results in an environmental quenching efficiency given by
\begin{equation}\label{eq_emh}
\varepsilon(m, M | M_0)=\frac{1}{c-\log M_0}\log\frac{M}{M_0}\,.
\end{equation}
We first use Equation (\ref{eq_emh}) with $\log(M_0)=13.25$ to fit the data
points shown in Figure \ref{fig_emh}. The best fit gives $c=16.37\pm0.06$ and is shown as the grey line in
Figure \ref{fig_emh}. We then use Equation (\ref{eq_fq4}) to fit the quenched fraction for
the total population shown in Figure \ref{fig_qfmh}, with $c$ fixed to $16.37$.
The best fit values for the other three parameters are $a=-0.2\pm0.004$,
$m_{\star}=6.5\pm0.2\times10^{10}\msunt$ and $d=0.61\pm0.02$.

\begin{figure}
\centering
\includegraphics[width=0.5\textwidth]{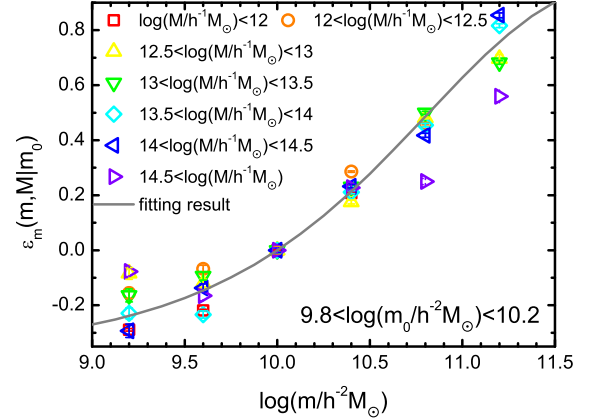}
\caption{The symbols with error bars show the mass quenching
efficiency $\varepsilon_{\rm m}(m, M | m_0)$ [see Equation (\ref{eq_hme})] as a function of
stellar mass $m$ for different halo masses, as indicated in the panel.
The efficiency is calculated by using galaxies with
$9.8\leq\log{m_0/\msunt}< 10.2$ to estimate $F_{\rm q}(m_0,M)$. The gray line is the fitting
curve (fitting formula Equation \ref{eq_hme2}). }
\label{fig_em}
\end{figure}

The results for $\varepsilon_{\rm m}(m, M | m_0)$ are presented in Figure \ref{fig_em}.
Here we choose $9.8\leq\log{m_0}< 10.2$ so that we have robust estimation of
$F_{\rm q}(m_0,M)$ for all the halo-mass bins. As one can see,
the $\varepsilon_{\rm m}(m, M | m_0)$ - $m$ relation is quite independent
of halo mass in all the halo mass bins except the most massive halo bin.
Since the quenched fraction can be
well fitted by Equation (\ref{eq_fq4}), $\varepsilon_{\rm m}(m, M | m_0)$ can be described by
\begin{equation}\label{eq_hme2}
\varepsilon_{\rm m}(m, M | m_0)
\equiv 1-\exp\left[-\left(\frac{m}{m_{\star}}\right)^d
+\left(\frac{m_0}{m_{\star}}\right)^d\right]\,.
\end{equation}
The prediction of this equation with $m_{\star}=6.5\times10^{10}\msunt$, $d=0.61$ and $m_0=10^{10}\msunt$
is plotted as the grey line in Fig.\,\ref{fig_em}.

\begin{figure}
\centering
\includegraphics[width=0.5\textwidth]{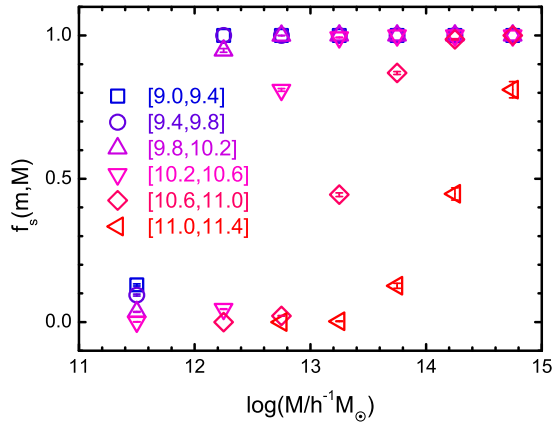}
\caption{The symbols show satellite fraction as a function of halo mass in various stellar mass bins.}
\label{fig_fsmh}
\end{figure}

\subsection{Quenching efficiencies for centrals and satellites}\label{ssec_ecs2}

As shown in Sections~\ref{ssec_ecs} and \ref{ssec_csp}, the density-based quenching
efficiencies for central and satellite populations depend on stellar mass in an opposite way,
and they counterbalance each other to produce a $m$-independent efficiency for
the total population. It is thus also interesting to examine the quenching efficiencies for
centrals and satellites separately by using halo mass, instead of the
density, as the environmental parameter. However, we will not repeat the same
analyses as in Section \ref{ssec_ecs} and \ref{ssec_csp},
for the following two reasons. First, the
stellar masses of centrals are strongly correlated with the halo masses of their
host groups, so that it is difficult to find a single halo mass
bin to define the environmental zero point for central galaxies of different masses.
Second, for satellites with given $(m,M)$, it is difficult to select
a large number of centrals of the same $(m,M)$ to form a control sample to
calculate the satellite-specific quenching efficiency. Because of these
we adopt a different approach, as described below.

To start with, we look at the satellite fraction as a function of
halo mass in different stellar mass bins, as shown in Figure \ref{fig_fsmh}.
Unlike the smooth relation between the satellite fraction with $\Delta$ shown in Figure \ref{fig_fsd},
$f_{\rm s}(m, M)$ as a function of $M$ resembles roughly
a step function, particularly for low-mass  galaxies.
The fraction is close to zero at $M<M_{\rm tr}$ and about one at $M>M_{\rm tr}$,
where $M_{\rm tr}$ is the mass scale at the transition of the step function and increases with
increasing stellar mass. Therefore, centrals and satellites dominate the galaxy population in
different regions in the $(m,M)$ plane and the two populations are comparable in number
only in a narrow region in the $(m,M)$ plane. This property of the satellite
fraction can be used to understand the quenching efficiency of centrals and
satellites over a large range in $m$ and $M$.

As shown in Figure \ref{fig_emh}, $\varepsilon(m, M | M_0)$ for galaxies of a
given $m$ follows the same correlation with $M$ below and above
$M_{\rm tr}(m)$ (marked by the vertical lines in the figure),
where galaxies are dominated by centrals and satellites, respectively.
This indicates that the quenched fractions in both the central
and satellite populations depend on the host halo mass in
a similar way not only in the $M$ range where the two populations
overlap, but also over the whole $M$ range. Similar analysis can also
be performed on the basis of the mass quenching efficiency $\varepsilon_{\rm m}$. For example, the four
stellar mass bins in $9.0<\log(m)<10.6$ are all dominated by centrals
in halos with $\log(M)<12$, but by satellites at $\log(M)>13$.
However, the mass quenching efficiency follows the same trend with
stellar mass in different halo mass bins, no matter whether the galaxies in
the halo mass bin are dominated by centrals or by satellites.
All these suggest that, whatever the quenching processes are,
they tend to produce a quenching efficiency that depends on halo
and stellar masses in a similar way for both centrals and satellites.
The results also suggest that the similarity in quenching efficiency
between centrals and satellites exists not only in the region of the
$(m,M)$ plane where the two populations overlap, but also over
the whole range of $(m,M)$ covered by the sample.

\begin{figure*}
\centering
\includegraphics[width=0.8\textwidth]{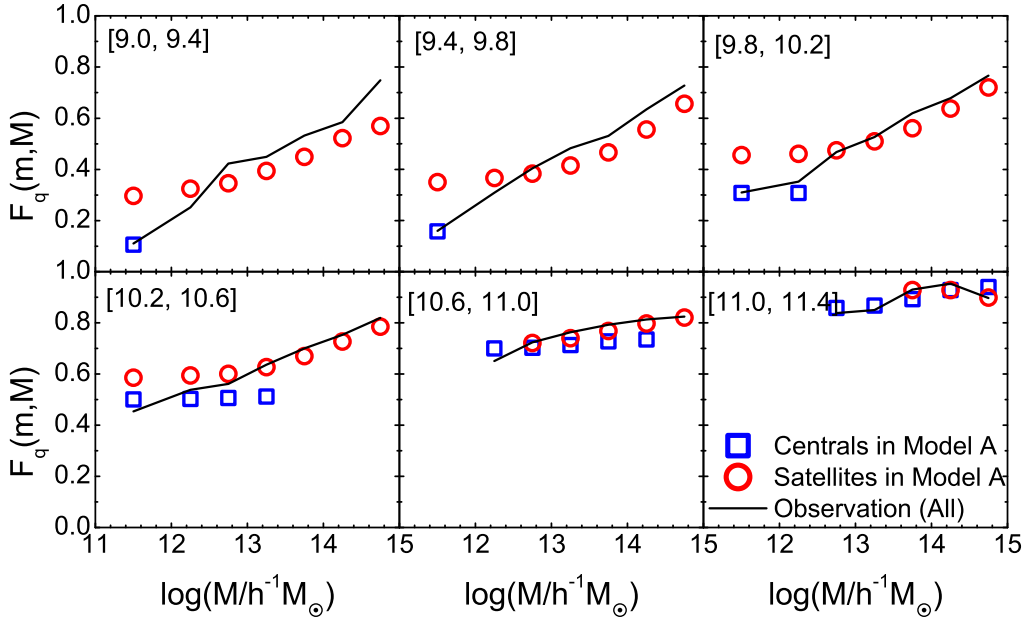}
\caption{Similar to Figure \ref{fig_qfmh} but for Model A.
The model results for the central (squares) and satellite (circles)
populations are plotted separately. The black lines show the
observational results for the total population.}
\label{fig_modela}
\end{figure*}

\section{Halo mass and assembly drive environmental quenching}
\label{sec_driver}

As we have shown above, the quenched fraction of galaxies depends on both
halo mass and environmental density. It is therefore important to examine
whether it is the halo mass or the environmental density that plays the
dominating role. To address this question, we construct two simple models,
in which the quenched fraction is assumed to be determined by the stellar mass
combined with one of the two environmental quantities.
We also try to understand whether assembly histories of dark matter
halos affect the galaxy properties in a third model. These models will also help
us to understand the apparent discrepancy between the results based on environmental
density and halo mass. Namely why is the density-based quenching efficiency
independent of the stellar mass for the total population but stellar-mass
dependent when centrals and satellites are analyzed separately
(Section \ref{sec_deqe}), while centrals and satellites follow
very similar trends in the halo-based quenching efficiency (Section \ref{sec_heqe})?

In the first model (hereafter Model A), the environmental density,
$\Delta$, is assumed to be the primary driver of the environmental
dependence of quenching. For each real galaxy in our SDSS sample, we
construct a corresponding model galaxy, which has exactly the same
$m$, $M$, $\Delta$, and the same identification as either a central or
a satellite. We assign a given model galaxy $i$ a quenching probability,
$q_i=F_{\rm q}(m, \Delta)$, according to its position in the $(m, \Delta)$
space (Figure \ref{fig_qfd4}). Note that centrals are treated differently
from satellites, because the observed $F_{\rm q,c}(m, \Delta)$ is very
different from $F_{\rm q,s}(m, \Delta)$.
We then use Equation (\ref{eq_fq3}) to estimate the average quenched
fraction for any given subset of the model galaxies. Figure \ref{fig_modela}
shows the predicted quenched fraction as a function of halo mass, $M$, for
galaxies in different $m$ bins.

As one can see, Model A predicts a positive dependence of the quenched
fraction on $M$, in rough agreement with the observation. This dependence
is expected from the fact that dark matter halos are tracers of the matter
density field. Note, however, that there is a marked difference
between the model prediction and the observational data. First of all, the model
predicts no significant dependence of $F_{\rm q,c}$ on $M$, contrary to the data which reveals a clear trend of increasing
$F_{\rm q,c}$ with increasing $M$ (cf., Figure \ref{fig_qfmh}).
Second, the predicted $M$-dependence for satellite galaxies is also
weaker than observed.  This discrepancy is particularly large for low
mass galaxies in low mass halos, where the model overpredicts $F_{\rm q,s}$
by as much as $0.2$. Finally, the model predicts significant differences
between the quenched fractions of centrals and satellites of similar $m$
and $M$, while such differences are absent in the observational data
(cf., Figure \ref{fig_qfmh}).

\begin{figure*}
\centering
\includegraphics[width=0.9\textwidth]{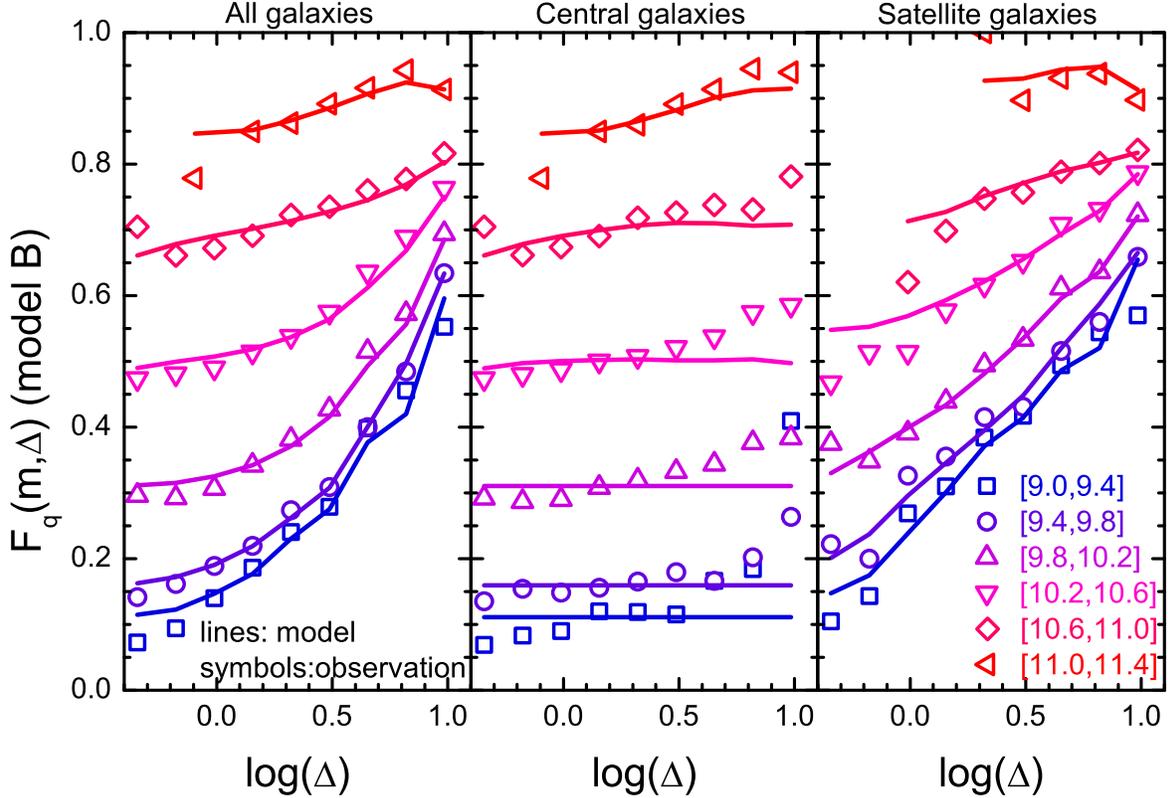}
\caption{Similar to Figure \ref{fig_qfd4} but obtained from Model B. The model results are shown in lines. The symbols show
the corresponding observational results (exactly the same as that in Figure \ref{fig_qfd4}).}
\label{fig_qfd4mb}
\end{figure*}

In the second model (hereafter Model B), halo mass instead of environmental
density is assumed to be the primary driver of the observed environmental
dependence of quenching. To test this hypothesis, we construct a model galaxy
sample by assigning each real galaxy in our SDSS sample a quenching probability
$q_i = F_{\rm q}(m, M)$ based on its position in the $(m, M)$ plane
(the black solid lines in Figure \ref{fig_qfmh}). Since the dependence of
$F_{\rm q}$ on $M$ and $m$ is very similar between
centrals and satellites, we do not distinguish between them when assigning $q_i$. Hence, in this model the probability for
a galaxy to be quenched is solely determined by the galaxy's stellar mass and
host halo mass, with centrals and satellite being treated in exactly the same
way.

\begin{figure*}
\centering
\includegraphics[width=0.8\textwidth]{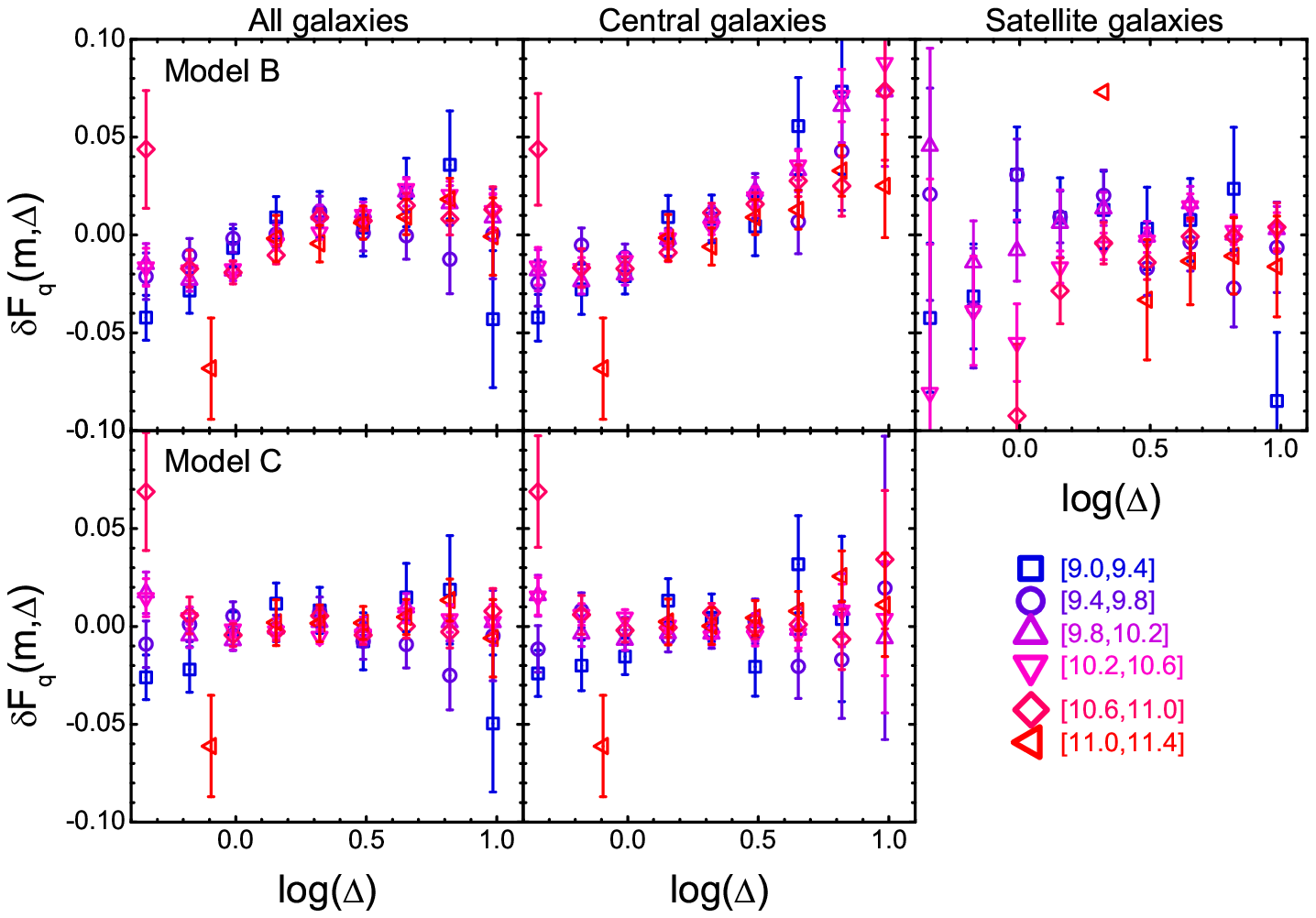}
\caption{The difference in the quenched fraction between the observation results
and model predictions (upper panels: Model B; lower panels: Model C). The results for satellites
are the same between the two models. The uncertainties are the error bars from the observational
data, as shown in Figure \ref{fig_qfd4}. The uncertainties in models are not taken into account.}
\label{fig_difq}
\end{figure*}

The quenched fractions predicted by Model B as a function of $\Delta$ and
$m$ [computed using Equation (\ref{eq_fq3})] are shown in Figure \ref{fig_qfd4mb}. The results closely
resemble those for the real galaxies. To better illustrate the quality
of the model, Figure \ref{fig_difq} plots the differences between the
model prediction and the data, $\delta F_{\rm q}(m,\Delta)$. Note that
the differences are fairly small, typically less than $\sim 0.05$ for
the entire galaxy sample. For centrals the discrepancies are slightly larger,
while for satellites the differences are comparable to the observational
uncertainties. Hence, Model B provides a fairly accurate description of the
data.  Taken together, the results from Models A and B strongly suggest that
the mass of the host halo is a far more important environmental parameter
for regulating quenching than is the matter density.

Before moving on, we try to understand some of the general trends predicted by Model B.
The resulting dependence of $F_{\rm q}$ on $\Delta$ is very different for centrals and satellites,
although the two populations are assumed to have exactly the same dependence of $F_{\rm q}$ on $m$ and $M$.
This arises because centrals and satellites of a given
$m$ cover very different ranges in $M$. Centrals with $9\leq\log (m)\leq 10.2$ usually reside
in halos with $\log (M) \sim 12$. The halo bias at this mass scale
is close to unity with little dependence on $M$ (see Sheth, Mo \& Tormen 2001).
Hence, the mean halo mass in the low-density regions is not very
different from that in the high-density regions, which explains
why $F_{\rm q,c}$ is almost independent of $\Delta$ for galaxies in this
$m$ range. As the mass of the central galaxy increases,
so does the mass of its host halo, which pushes it into the regime where
halo bias is larger than unity and has a strong dependence on halo mass.
Consequently, halos in high density regions are, on average, more
massive than those in low density regions, even if they contain centrals of the
same stellar mass. This explains why the dependence of $F_{\rm q,c}$ on $\Delta$
becomes stronger for centrals with higher stellar masses, as shown in
Figure \ref{fig_qfd4mb}. However, for a given $m$, the distribution
in $M$ is quite narrow (see Yang et al. 2009), so the $\Delta$-dependence predicted by Model B remains weak.
For satellites, the situation is very different. At the low $m$ end, satellites reside in halos that
cover a very wide range in $M$, which gives rise to a very strong dependence on $\Delta$. As $m$
increases, the dispersion in $M$ decreases, which weakens the dependence of $F_{\rm q,s}$ on $\Delta$.

As evident from Figure \ref{fig_difq}, Model B slightly overestimates $F_{\rm q,c}$ at
the low-$\Delta$ end while underestimating it at the high-$\Delta$ end, a trend that is evident
for every stellar mass bin. It suggests that the quenching of (central) galaxies depends not only on halo
mass and stellar mass, but also on some other halo properties that are correlated with
the environmental density. One such halo property is halo assembly history, which is correlated
with the environmental density (known as assembly bias).
In order to see if the discrepancy can be explained by halo assembly bias,
we need to know, for each individual group, the formation redshift (hereafter
$z_{\rm f}$) that characterizes its assembly history. Our ELUCID simulation
is a constrained simulation in the SDSS DR7 region and can reliably reproduce
most of massive groups (see paper III), and here we make use of the information
it provides to estimate the formation redshifts for individual groups. To do
this, for each group, we search all halos in the simulation that have mass
differences less than 0.3 dex with, and distance less than $5\mpc$ to, the
group in question.  Most of the groups ($\sim97.4\%$) have at least
one halo companion defined in this way, and we assign the formation redshift
of the nearest halo to the group. The formation redshift is defined as the highest
redshift at which half of the final halo mass has assembled into progenitors more
massive than $10^{11.5}\msun$ (see Neistein, van den Bosch \& Dekel 2006; Li et al. 2008).
The choice of this mass limit is motivated by the fact that it corresponds to the
halo mass at which the star formation efficiency is the highest at different
redshifts (see Lim et al. 2017). For groups with $\log(M)<12.0$
that do not have accurate halo mass estimates in the
group catalog, we only search for halo companions with
$11.7<\log(M)<12.0$, where the lower mass limit
($11.7$) is adopted so that all halos have reliable estimates of
halo formation redshifts in the ELUCID simulation. We have also used
other definitions of halo formation redshift, such as the
redshift at which the main progenitor reaches half of the present-day
halo mass. The results are very similar.

\begin{figure*}
\centering
\includegraphics[width=0.9\textwidth]{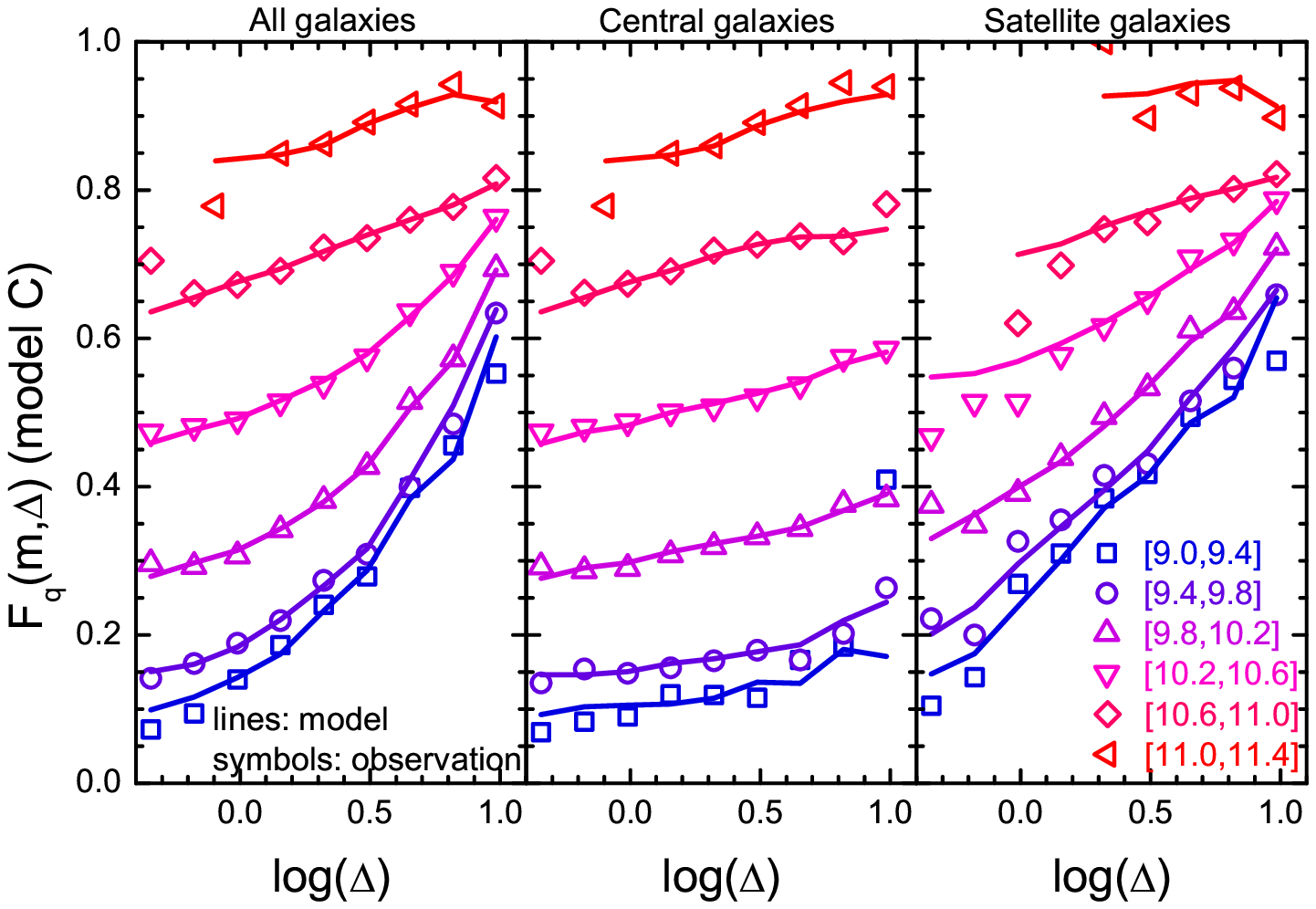}
\caption{Similar to Figure \ref{fig_qfd4} but obtained from Model C. The model results are shown in lines. The symbols show
the corresponding observational results (exactly the same as that in Figure \ref{fig_qfd4}).}
\label{fig_qfd4mc}
\end{figure*}

A third model, Model C, is then constructed on top of Model B.
In Model C, the quenching probability of a model galaxy depends not only
on $m$ and $M$, but also on the formation redshift of its halo.
The simplest way to link $z_{\rm f}$ to quenching probability is to assume that
a galaxy is quenched when $z_{\rm f}>z_{\rm th}$, where $z_{\rm th}$ is a
formation redshift threshold. In reality, however,  galaxies with given
$z_{\rm f}$, $m$ and $M$, must have some dispersion in their star formation rate.
To mimic this, we introduce a dispersion in $z_{\rm f}$ for each system before
applying the criterion $z_{\rm f}>z_{\rm th}$ to select the quenched fraction.
In practice, for a galaxy $i$ with $z_{\rm f}=z^i_{\rm f}$, we use a Monte Carlo
method to generate 500 mock galaxies, with their formation redshifts
($z^i_{\rm f,m}$) randomly drawn from a Gaussian distribution with
the mean value equal to $z^i_{\rm f}$ and a width $\sigma_{\rm z}$.
Then, for a given $z_{\rm th}$, the quenching probability of the model
galaxy, $q_i$, is set to be the fraction of the mock galaxies that have
$z^i_{\rm f,m}> z_{\rm th}$ among the 500 mock galaxies. In order to introduce
the dependence on $m$ and $M$, the threshold $z_{\rm th}$ is required
to be a function of the two quantities, and is determined by the criterion that
the dependence on $m$ and $M$ for the model galaxies is
exactly the same as that for real galaxies. Note that in our model
the formation redshift dependence is only considered for central galaxies;
satellites are treated in exactly the same way as in Model B.

When $\sigma_{\rm z}$ is set to be 0, the dependence of
$F_{\rm q,c}(m, \Delta)$ on $\Delta$ for the model galaxies is found to be much stronger than that
for real galaxies. We have experimented a series of values for $\sigma_{\rm z}$, and found
that the model matches the observation the best when $\sigma_{\rm z}\sim 0.8$.
Figure \ref{fig_qfd4mc} shows the quenched fraction as a function of $m$ and
$\Delta$ obtained from this best model. A significant increasing trend
of the quenched fraction with $\Delta$ is now produced for central galaxies in most
of $m$ bins, as is seen in the observational data.
The lower two panels of Figure \ref{fig_difq} show the difference between the
observational quenched fraction and the predictions of Model C for the total
population and centrals, respectively. In contrast to Model B, the dependence of
$\delta F_{\rm q}(m,\Delta)$ on $\Delta$ almost completely disappear in Model C.

\begin{figure*}
\centering
\includegraphics[width=0.9\textwidth]{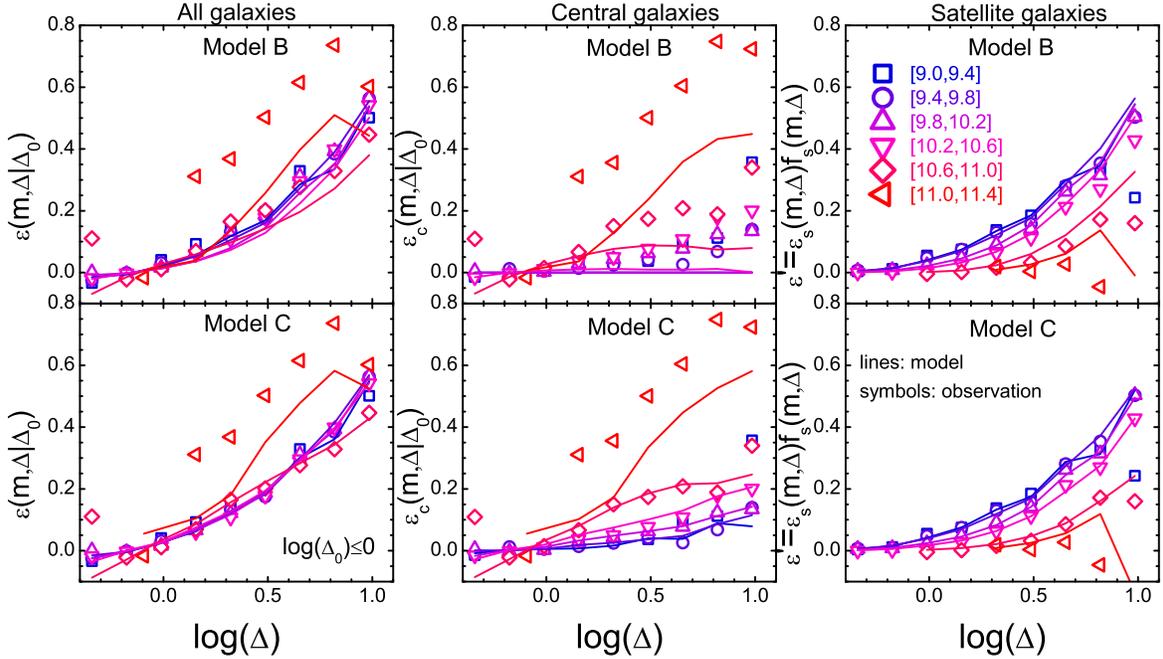}
\caption{The environmental quenching efficiencies for model B (upper panels)
and model C (lower panels). Lines show the model prediction and symbols show the corresponding observational data.
Left panels: efficiency of Equation (\ref{eq_dee1}). Middle panels: efficiency of Equation (\ref{eq_dec}).
Right panels: efficiency of Equation (\ref{eq_dee3}). For the most massive
galaxies in Model C and in the observational data, the data point at
$\Delta\sim 1.0$ lies outside the figure boundary.}
\label{fig_emodel}
\end{figure*}

We then compute the environmental quenching efficiency, defined in Equation (\ref{eq_dee1})
with the zero point estimated from data at $\log\Delta\leq 0$, for Model B and C.
The results are presented in the left panels of Figure \ref{fig_emodel}.
The quenching efficiency predicted by Model B exhibits a very weak dependence on $m$.
This is not surprising since the halo-based environmental efficiency,
${\varepsilon_{\rm c}}(m, M | M_0)$, is independent of stellar mass (Figure \ref{fig_emh})
and since halo mass is the only driver of the environmental quenching in Model B.
There is a small deviation between the model prediction and the observation for most
of the stellar mass bins. Such deviation is absent in Model C, suggesting that it has the same
origin as that shown in Figure \ref{fig_difq}.

We also show the predictions of Model B and C
for ${\varepsilon_{\rm c}}(m, \Delta | \Delta_0)$, defined by Equation (\ref{eq_dec}),
and for $\varepsilon'(m, \Delta | \Delta_0)$, defined by Equation (\ref{eq_dee3}),
in the middle and right panels of Figure \ref{fig_emodel}. In Model B, one can see a clear trend
that, at a fixed $\Delta$, ${\varepsilon_{\rm c}}$ increases,
while $\varepsilon'$ decreases, with $m$. As discussed above,
the halo mass distribution becomes broader for centrals but narrower for satellites as stellar
mass increases, which explains the opposite trends in the quenched efficiency as a function of
stellar mass for centrals and satellites. After taking into account the assembly bias
effect (Model C), the dependencies on stellar mass for both centrals and satellites are enhanced
and the predictions resemble more closely the observational results.

Moreover, the predicted efficiency for the most
massive galaxies is now closer to the other mass bins than the observational result, giving
further support to the hypothesis that the discrepancy found for the most massive
galaxies in the observational data is mainly caused by the uncertainty in the estimation of $F_{\rm q}(m,\Delta_0)$. The uncertainty
is largely eliminated in the model predictions, because the value of $q_i$ assigned to
galaxies in the low-density regions is the average over a larger sample of galaxies.

\section{The quenching of centrals is not special}\label{sec_cen}

As shown in Section \ref{ssec_qfh} and \ref{ssec_ecs2}, we find that centrals and satellites
follow the same correlation of quenched fraction with stellar mass and host
halo mass. This result appears to be in conflict with
those obtained by some earlier studies (e.g. van den Bosch et al. 2008;
Wetzel et al. 2012; Peng et al. 2012), where quenching is found to depend
strongly on whether a galaxy is a central or a satellite. However,
as discussed above, host halo mass ranges
covered by centrals and satellites of the same stellar mass are very different.
Thus, if halo mass is not used as a control parameter when comparing the two
populations, as is the case in most of the previous studies,
then one is comparing centrals in low mass halos with satellites in massive ones
and neglecting the strong dependence on halo mass.
We believe that this is the origin of the discrepancy and that there is
no conflict between our results and those obtained in these earlier
studies.

It is possible that the similarity between centrals and satellites
we find in the data is not real, but caused by errors in the group finder
used to identify groups and central galaxies in them. As one can see
from Figure \ref{fig_fsmh}, the satellite fraction, $f_{\rm s}(m,M)$, resembles
roughly a step function and is
close to zero at $M<M_{\rm tr}$  and about one at $M>M_{\rm tr}$ ($M_{\rm tr}$
is the mass scale at the transition of the step function).
Thus, even if the mis-identification fraction is small in the whole population
(see Hirschmann et al. 2014), the galaxies that are identified as satellites
at $M<M_{\rm tr}$ may be significantly contaminated by centrals. In this case, the
similarity between centrals and `satellites' of a given stellar mass in their
quenched fraction may be produced by the false identification
between centrals and satellites, rather than a real similarity.

However, it is difficult to explain the
$m$-independence of the environmental quenching efficiency and the
$M$-independence of the stellar mass quenching efficiency
(Section \ref{sec_heqe}) by central/satellite mis-identifications
alone. If centrals and satellites had quenching properties
that depend on the halo and stellar masses in significantly
different ways, the environmental quenching efficiency, $\varepsilon(m, M | M_0)$,
would be expected to correlate with the halo mass in different
ways depending on whether $M<M_{\rm tr}$ or $M>M_{\rm tr}$, as the galaxy
populations in the two mass ranges are dominated by centrals and satellites,
respectively (see Fig.~\ref{fig_fsmh}). Moreover, since $M_{\rm tr}$
increases with increasing stellar mass, $\varepsilon(m, M | M_0)$ would also
be expected to vary with stellar mass. Similarly, it would also lead to
$M$-dependence in the mass quenching efficiency, $\varepsilon_{\rm m}(m, M | m_0)$.
Both are inconsistent with the observational results.
We thus conclude that the similar quenching properties of centrals
and satellites are not produced by mis-identifications
between the two populations, and are real within the statistical
uncertainties of the data.

It should be pointed out that this interpretation is based on the
premise that the inferred halo mass is sufficiently accurate. Recently,
Campbell et al (2015) used a mock catalog including galaxy
color to check the color-dependent statistics inferred from group catalogs.
They compared the marginalized dependence of the quenched (red) fraction
on halo mass obtained from their mock group catalog with
that obtained directly from the
simulation used to construct the mock catalog, and found that group finders
tend to reduce the difference in the quenched fraction between centrals
and satellites (see their figure 13). This systematic error is
produced by the combined effect of group membership determination,
central/satellite designation, and halo mass assignments.
If the real difference between centrals and satellites is small,
it may be washed out by this error. However, we want to point out that
if this systematic error shown in the mock group catalog in Campbell et al. indeed
affects the SDSS group catalog in a similar way, one would expect to see that
the marginalized dependence of the quenched fraction
on halo mass is similar between centrals and satellites for the SDSS group catalog.
It is apparently inconsistent with what we found (the middle panel of Figure \ref{fig_qf}).
This suggests that the error caused by group finder depends on the galaxy formation model that is used to
construct the mock catalog. Unfortunately this also means that
the impact of the inaccuracy of the group finder on the results
obtained here from the SDSS group catalog is unclear.

\begin{figure*}
\centering
\includegraphics[width=0.8\textwidth]{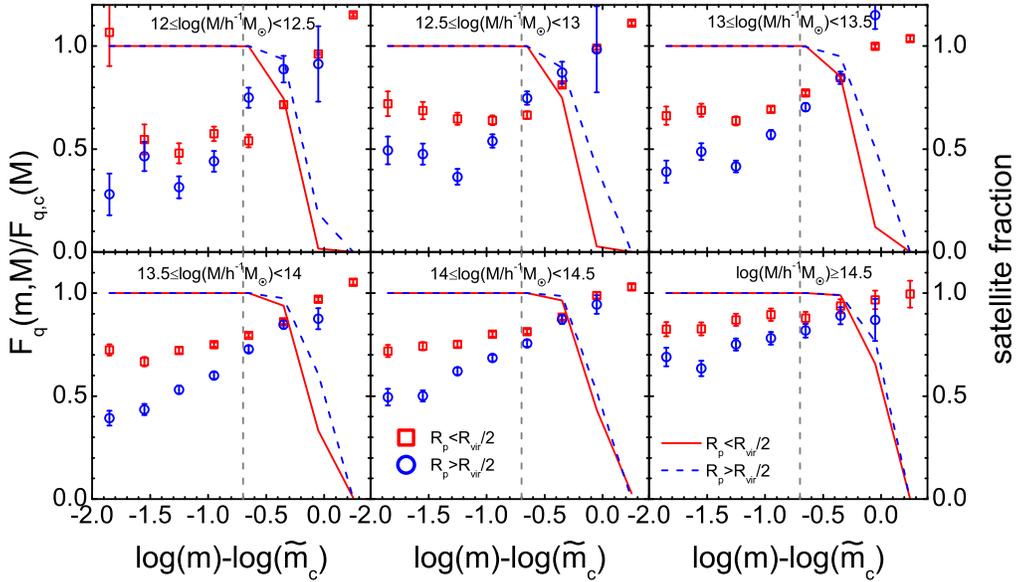}
\caption{The symbols show the scaled quenched fraction as a function of the scaled
stellar mass for galaxies with $R_{\rm p}<R_{\rm vir}/2$ (red squares) and
$R_{\rm p}>R_{\rm vir}/2$ (blue circles) in six halo mass bins. The red and blue lines show the
satellite fraction as a function of scaled stellar mass. The vertical dashed lines indicate $\log(m)- \log(\tilde{m}_{\rm c})=-0.7$. Here the quenched fraction,
$F_{\rm q}(m,M)$, is scaled by $F_{\rm q,c}(M)$, the quenched fraction of central
galaxies in the corresponding halo mass bins, and the stellar mass is scaled with
$\tilde{m}_{\rm c}$, the median of the stellar mass
of central galaxies.}
\label{fig_qfqfc}
\end{figure*}

One way to bypass the uncertainty in the central versus satellite
identification is to study the dependence of galaxy quenching on
the locations of galaxies in their host halos. Instead of looking
at centrals versus satellites, we can look at galaxies at different
locations in a halo. By definition, centrals belong to the
innermost population. Previous studies (e.g. Wetzel et al. 2012; Woo et al. 2013; Bluck et al. 2016)
found that the quenched fraction increases with decreasing distance to
the halo center (halo-centric radius). In contrast, the
similarities between centrals and satellites found in this paper
seem to suggest that quenching is quite independent of the location within
halos. To find the cause of this discrepancy, we divide the total galaxy population
into two: the inner and outer sub-populations, according to whether or not their projected distances to the halo centers
($R_{\rm p}$) are smaller or larger than half of the virial radii ($R_{\rm vir}$).
Here the center of a halo is the luminosity-weighted position of its member galaxies.
Figure \ref{fig_qfqfc} shows the quenched fraction as a function
of stellar mass for the two sub-populations in the six halo mass bins used above. To compare with the central galaxies in the corresponding
halo mass bin, the quenched fraction is scaled with the quenched fraction of centrals and the stellar mass is also scaled with the median of the stellar mass of central galaxies ($\tilde{m}_{\rm c}$). For comparison, the satellite fractions in the two
sub-populations are also shown in the figure.

For small galaxies, the quenched fraction clearly increases with the decrease of halo-centric radius,
consistent with previous studies, while for massive galaxies the dependence on
halo-centric radius is very weak. The characteristic stellar mass above which the $R_p$ dependence becomes
weak is $\log m\simeq \log(\tilde{m}_{\rm c})-0.7$ (indicated by the vertical lines) and almost independent of
halo mass. This clearly demonstrates that the quenching probability is independent of its location in the halo,
as long as a galaxy has a stellar mass larger than about one fifth of the
mass of its central galaxy. For galaxies with $\log m\sim \log \tilde{m}_{\rm c}$, the
scaled quenched fraction is close to unity for both the inner and outer
populations. At this stellar mass, the satellite fraction is close to zero
for halos with $\log(M)<12.5$ and increases to
$>60$\% for $\log(M)>14.5$. It indicates that, when galaxies have
stellar masses comparable to those of their centrals, the quenched fraction
is the same as that of their centrals, no matter where they are located and
whether they are identified as centrals or not. Alternative separations of inner and outer
regions at $0.3R_{\rm vir}$ and $0.4R_{\rm vir}$ have no significant impact on this conclusion.
These demonstrate again that centrals are not special as far as
their quenching properties are concerned. Our results are also not in conflict
with previous finding, as a strong dependence on halo-centric radius
is present but only for galaxies that are much less massive than the
centrals in their corresponding halos.

\section{Summary and Discussion}
\label{sec_summary}

In this paper, we present a detailed investigation about the environmental dependence of
quenching of star formation using a large sample of galaxies constructed from
the SDSS. We adopt two quantities to describe the different aspects of galaxy
environments: the environmental mass densities, smoothed on a scale
of $4\mpc$ (half width of Guanssian kernel) at the positions of individual galaxies,
and the masses of the host halos within which galaxies reside.
The mass densities are obtained from the ELUCID simulation, a constrained
$N$-body simulation in the SDSS volume,  while the halo masses are based on a galaxy
group catalog constructed with a halo-based group finder. Our main findings
are summarized as follows.

\begin{itemize}
\item The quenched fraction of galaxies increases systematically with galaxy stellar mass,
environmental density and host halo mass. When analyzed separately,
centrals and satellites show very different density-dependence:
while the dependence is strong for satellites, it is
weak or even absent for central galaxies. The environmental
effect is stronger for centrals with higher stellar masses,
while satellites show the opposite trend.
In contrast, the dependence of the quenched fraction on halo mass is almost
the same for both centrals and satellites, although the two populations cover different ranges of halo masses.

\item
For the total galaxy population, the quenching efficiency, defined
as the quenched fraction of galaxies in a given environment relative to
a zero-point population, is found to be almost independent of galaxy stellar mass over a wide
mass range, $9\leq\log (m)\leq11.4$), no matter which
environmental parameter (mass density or halo mass) is adopted.
This suggests that the strong stellar mass-dependence of the quenched
fraction is predominantly produced by such a dependence of the zero point.

\item
When central and satellite galaxies are analyzed separately,
the density-based quenching efficiency is found to increase systematically with stellar mass
for centrals but to decrease for satellites, and the stellar mass dependence
is stronger for galaxies of higher stellar masses. The opposite trends
seen in centrals and satellites compensate each other so as to
make an almost stellar mass-independent quenching efficiency for the total
population described above. The results thus do not support the hypothesis
proposed in previous studies that the mass independence of the quenching efficiency is
only due to satellite quenching.

\item
Centrals and satellites are found to follow the same trends in both the halo mass
and stellar mass based quenching efficiencies, contrary to the
efficiency defined by the environmental density. It indicates that quenching
does not depend strongly on the location within halo,
at least for galaxies with stellar masses comparable to those of the centrals.
Further investigation of the dependence of the quenched fraction on
the distance to halo center shows that the distance dependence is only
important for galaxies with stellar masses that are lower than one fifth of
the masses of their centrals but insignificant for more massive galaxies.

\item
A model, in which the quenching probability of a galaxy is assumed to be
determined by the galaxy stellar mass combined with the host halo mass (Model B),
can well reproduce the observed dependence of the quenched fraction on environmental
density for the total population, as well as separately for centrals and satellites.
In contrast,  a model in which the quenching probability is assumed to
be determined by galaxy mass and environmental density (Model A)
predicts too weak dependence on halo mass and different halo-mass dependence
between centrals and satellites, in conflict with observation.
These suggest that halo properties are the driver of the environmental quenching
seen in the observational data.

\item
Model B is found to slightly overestimate the quenched fraction in
the low density end and to underestimate it at the high density end for central galaxies.
It suggests that galaxy quenching depends not only on halo mass
but also on some other halo properties that are correlated with the environmental density.
A model (Model C), which takes into account halo assembly bias,
can explain the discrepancy between Model B and the observational results.

\item
The environmental quenching efficiency based on the mass density
predicted by both Model B and Model C (where the halo-mass is the primary driver of the environmental effect) is found to be
independent of stellar mass, consistent with observational results. 
This strongly suggests that the stellar mass independence of the density-based efficiency
originates from the stellar mass independence of the halo-based efficiency.

\item
The difference in star formation quenching between centrals and satellites
found in this paper and in numerous previous studies are mainly due to the
difference in the host halo mass ranges covered by the two populations,
and not produced by the difference in the correlation of quenching
probability with halo mass between the two populations.
\end{itemize}

Many mechanisms of quenching star formation in galaxies have been proposed in the literature,
such as virial shock heating to accretion flows, the feedback from active galactic nuclei and supernova,
and the stripping of hot and cold gas associated with galaxies (see e.g. Gabor et al. 2010 for a more comprehensive discussion).
The strengths of these quenching mechanisms depend on stellar mass and halo mass in different ways.
Furthermore, because of the special positions assumed for central galaxies in their
halos, some of the processes may have different impact between centrals and
satellites. In what follows we discuss how our findings can be used to
constrain these different quenching mechanisms.

AGN feedback has been proposed to quench  cooling flows in massive halos and suppress
star formation in massive galaxies (e.g. Croton et al. 2006; Bower et al. 2006; Cui, Borgani, \& Murante 2014). The strength of AGN
feedback is usually assumed to be more efficient for more massive galaxies that contain more
massive black holes in more massive halos. AGN feedback may thus be able to produce
the positive dependency of the quenched fraction and quenching efficiency on
galaxy stellar mass and halo mass found in this paper.  Moreover, AGN feedback may affect
both centrals and satellites in a similar way, again consistent with our observation.
Numerous observations have provided evidence for AGN feedback through radio jet and massive
outflows driven by radiation pressure (e.g. Best et al. 2007; Wang et al. 2011b; Fabian et al. 2012).
However, the details how AGN feedback is coupled with the gas and regulates the
star formation in galaxies are still uncertain.

Hydrodynamical simulations (e.g. Kere{\v s} et al. 2005; 2009) have revealed that galaxies acquire
their baryonic mass primarily through cold gas flows along  filamentary structures around
halos, and such cold accretion can be heated and suppressed by virial shocks in the host halo (see also Dekel \& Birnboim 2006).
This process can result in a decrease in the cold accretion as a function of halo
mass, as radiative cooling is less effective in more massive halos
(Ocvirk, Pichon \& Teyssier 2008; Kere{\v s} et al. 2009). If the star formation in
galaxies is fuelled mainly by the cold accretion, an increasing trend of
the quenched fraction and the quenching efficiency with halo mass is expected.
Moreover, galaxies in simulations are found to continue to acquire cold gas
after becoming satellites, and the cold accretion is also affected
by the shock in the host halos (Kere{\v s} et al. 2009), which is consistent
with our finding that the quenching properties of centrals and satellites are similar.
Unfortunately, it is unclear how the suppression of cold accretion depends on
the stellar mass of a galaxy in a halo, and so it is unclear if this mechanism can
accommodate the dependence of quenching on stellar mass seen in the observation.

The stripping of gas from galaxies by ram-pressure in hot halo is expected to be more efficient for galaxies
of lower masses living in higher mass halos  (see Henriques et al. 2016 for discussion).
Therefore this mechanism predicts that quenching should be important only in massive halos
with hot halo gas, and that the quenched fraction and quenching efficiency should increase with
decreasing stellar mass.  These predictions seem to be at odds with the positive dependence of
the quenched fraction  (quenching efficiency) on stellar mass found in this paper, and with the fact
no characteristic halo mass is seen in the quenching fraction - halo mass relation.  In addition,  ram
pressure  stripping is expected to be important only for satellite galaxies that are orbiting
in halos, but unimportant for centrals that sit close to the bottoms of the halo gravitational potential wells.
If such stripping process dominates quenching of star formation,  then centrals and satellites are expected
to be affected differently.  This seems contrary to the results that the quenching efficiencies of centrals and
satellites are correlated with halo mass and galaxy mass in a similar way. Our results thus suggest that
ram pressure stripping alone cannot be the dominant quenching processes for the whole population.
Similar argument may also be made for tidal stripping. However, since the effect of the tidal stripping
is determined by the local mass density relative to the mass density of the galaxy to be stripped,
this effect may also be important in relatively low-mass halos, which may be in better
agreement with observation.

The dependence of quenching on halo-centric radius for low-mass galaxies suggests
that the significance of the underlying physical processes depends on the locations of
galaxies in their host halos. The satellite-specific processes, such as ram pressure stripping
and tidal stripping,  are expected to be more efficient near the halo center,  and so may
be able to produce the dependence on halo-centric radius observed in the data.
However, as discussed above, if these processes dominate the quenching of satellites
and do not operate on centrals,  then why do central galaxies have quenching properties similar
to satellite galaxies of the same stellar masses?  It may be that centrals are not
special, and these satellite-specific processes also operate on centrals.  This is in fact
consistent with the fact that centrals are not at rest within the halo potential wells, and
not even located at the halo center (e.g. Skibba et al. 2011).
Indeed, one of the main conclusions that can be drawn from our results
is that the central in a halo is not special, as far as its star formation
quenching is concerned. Peng et al. (2010) found that the environmental quenching and mass quenching can be well
separated from each other for galaxies at $z\sim 1$. It suggests that such a
conclusion also holds for these galaxies.
However, as mentioned above, this similarity between centrals and satellites is not expected in
some galaxy formation models, where centrals and satellites are assumed to be
affected by different quenching processes.

The processes discussed above, which are all confined within halos, can not account for the residual
dependence on environmental density after removing the halo-mass effects. We thus include halo assembly history as
an additional parameter, which is known to be influenced by large scale environment (e.g. Gao et al. 2005).
Local tidal field is thought to play a key role in shaping the halo assembly bias (Wang et al. 2007;2011a; Hahn et al. 2009;
Shi et al. 2015; Paranjape et al. 2017; Borzyszkowski et al. 2017). For example, for a small halo in a high density region,
the material around the halo is accelerated by the local tidal field so that the halo growth is significantly suppressed.
These processes (on the scale much larger than the galactic scale) are unlikely to directly influence the star formation in galaxies. Therefore, a correlation between the star formation and the halo assembly history has to be introduced in order to explain
the residual dependence. It is worthwhile to note that it is unclear whether these large scale processes affect
baryonic gas and dark matter in the same way. If it is not, additional effect on star formation should be taken into account.

The discussions given above provide some qualitative assessments about
some of the quenching processes that have proposed in the literature,
in connection to the observational results we find in this paper. To
constrain theoretical models in a quantitative way, however,
detailed modeling of the various quenching processes, as well as
thorough analyses of all potentially important observational selection effects
are needed. In a forthcoming paper,  we will use mock catalogs constructed
from hydrodynamic simulations and semi-analytic models of galaxy formation
to compare galaxy formation models with the observational results
obtained here (E. Wang et al. in preparation).

\section*{Acknowledgments}

We thank an anonymous referee for a useful report. This work is supported by the 973 Program (2015CB857002), NSFC
(11522324, 11733004, 11421303, 11233005, 11621303),  and the Fundamental
Research Funds for the Central Universities. H.J.M.
would like to acknowledge the support of NSF AST-1517528 and NSFC-11673015.
S.H.C and Y.Y. are supported by the Fund for Fostering Talents in Basic
Science of the National Natural Science Foundation of China
NO.J1310021. FvdB is supported by the Klaus Tschira Foundation and by the US
National Science Foundation through grant AST 1516962. WC is supported by the {\it Ministerio de Econom\'ia y Competitividad} and the {\it Fondo Europeo de Desarrollo Regional} (MINECO/FEDER, UE) in Spain through grant AYA2015-63810-P as well as the Consolider-Ingenio 2010 Programme of the {\it Spanish Ministerio de Ciencia e Innovaci\'on} (MICINN) under grant MultiDark CSD2009-00064.
The work is also supported by the Supercomputer Center
of University of Science and Technology of China and the High Performance Computing
Resource in the Core Facility for Advanced Research Computing
at Shanghai Astronomical Observatory.

\end{document}